\documentclass[11pt,a4paper]{article}
\usepackage{jcappub}

\usepackage{bm}

\newcommand{\alt}{\mbox{\;\raisebox{.3ex}
  {$<$}$\!\!\!\!\!$\raisebox{-.9ex}{$\sim$}\;}}

\newcommand{\be}{\begin{equation}}
\newcommand{\ee}{\end{equation}}
\newcommand{\bea}{\begin{eqnarray}}
\newcommand{\eea}{\end{eqnarray}}

\renewcommand{\vec}[1]{{\bm #1}}
\begin{document}



\title{Efficient calculation of cosmological neutrino clustering with both linear and non-linear gravity}

\author[a]{Maria~Archidiacono, Steen~Hannestad}

\affiliation[a]{Department of Physics and Astronomy, University of Aarhus,
 DK-8000 Aarhus C, Denmark}

\emailAdd{archi@phys.au.dk, sth@phys.au.dk}

\abstract{We study in detail how neutrino perturbations can be followed in linear theory by using only terms up to $l=2$ in the Boltzmann hierarchy. We provide a new approximation to the third moment and demonstrate that the neutrino power spectrum can be calculated to a precision of better than $\sim$ 5\% for masses up to $\sim$ 1 eV. The matter and CMB power spectra can be calculated far more precisely and typically at least a factor of a few better than with existing approximations.
We then proceed to study how the neutrino power spectrum can be reliably calculated even in the presence of non-linear gravitational clustering by using the full non-linear gravitational potential derived from semi-analytic methods based on $N$-body simulations in the Boltzmann evolution hierarchy. Our results agree extremely well with results derived from $N$-body simulations.
}

\maketitle

\section{Introduction}

Within the next few years cosmological structure formation will be probed in greater detail than ever before by new and very large surveys, most notably EUCLID \cite{Laureijs:2011gra} and LSST \cite{lsst}.
In terms of observables such as the matter or shear power spectra the precision will be at the 1\% level (see e.g.\ \cite{Schneider:2015yka}). While this opens great possibilities for probing for example dark energy and the mass of neutrinos, it also puts very stringent requirements on theoretical calculations of those same observables.
Massive neutrinos are particularly challenging because they cannot simply be treated using the same particle representation of the underlying distribution function as is done for cold dark matter. The reason is that neutrinos are very light and thus have thermal velocities comparable to, or greater than, the gravitational streaming velocities.

Furthermore, massive neutrinos are much more difficult to follow in linear theory Boltzmann codes such as CAMB \cite{CAMB} and CLASS \cite{CLASS} than other particles because the free-streaming length is different for different neutrino momenta. Solving the full system of equations for a representative selection of neutrino modes is computationally demanding, particularly if the code has to be run many times such as in Markov Chain Monte Carlo applications.

In this work we first discuss a new and very precise approximation to the neutrino Boltzmann hierarchy in linear codes. Even though the neutrino Boltzmann hierarchy is truncated the neutrino power spectrum can be computed with a precision of a few percent, far better than existing approximation schemes.

We then proceed to describe an extremely efficient way of calculating the neutrino power spectrum in the regime of non-linear gravitational clustering by  keeping all calculations in Fourier space. We will demonstrate how to calculate the neutrino power spectrum using the full non-linear gravitational potential derived from semi-analytic fitting with HALOFIT~\cite{halofit, Takahashi:2012em, Viel2011}. We will show that the derived $P_\nu(k)$ is equivalent to what can be calculated in real space using the approach in Ref.~\cite{AliHaimoud:2012vj}.

The paper is organised as follows: In section 2 we discuss the linear evolution equations for massive neutrino perturbations as well as an extremely efficient way of solving them by means of truncation.
In section 3 we proceed to discuss how to follow neutrino perturbations in the linear regime but under the influence of the fully non-linear gravitational potential.
Finally, section 4 contains a summary and a discussion of our results.

\section{Linear theory evolution of neutrino perturbations}

The evolution of any given particle species in a dilute gas with no quantum entanglement can be followed using the single particle Boltzmann equation. For neutrinos in the late time universe interactions can be safely neglected and we can use the collisionless Boltzmann equation. The notation used here follows that of Ma and Bertschinger \cite{Ma} and uses the synchronous gauge (the equations could equally well be written in e.g.\ conformal gauge).
As the time variable we use conformal time, defined as $d \tau = dt/a(t)$, where $a(t)$ is the scale factor. Also, as the momentum variable we shall use the
comoving momentum $q_j \equiv a p_j$. We further parametrize $q_j$ as
$q_j = q n_j$, where $q$ is the magnitude of the comoving momentum and $n_j$ is a unit 3-vector specifying direction.

The collisionless Boltzmann equation can generically be written as
\begin{equation}
L[f] = \frac{Df}{D\tau} = 0,
\end{equation}
where $L[f]$ is the Liouville operator.

One can then write the distribution function as
\begin{equation}
f(x^i,q,n_j,\tau) = f_0(q) [1+\Psi(x^i,q,n_j,\tau)],
\end{equation}
where $f_0(q)$ is the unperturbed distribution function and $\Psi$ its perturbation.

In synchronous gauge the Boltzmann equation can be written as an evolution equation for $\Psi$ in $k$-space \cite{Ma}
\begin{equation}
\frac{1}{f_0} L[f] = \frac{\partial \Psi}{\partial \tau} + i \frac{q}{\epsilon}
\mu \Psi + \frac{d \ln f_0}{d \ln q} \left[\dot{\eta}-\frac{\dot{h}+6\dot{\eta}}
{2} \mu^2 \right] = \frac{1}{f_0} C[f],
\label{eq:boltzX}
\end{equation}
where $\mu \equiv n^j \hat{k}_j$ and $\epsilon = (q^2+a^2 m^2)^{1/2}$.
$h$ and $\eta$ are the metric perturbations, defined from the perturbed space-time
metric in synchronous gauge \cite{Ma}
\begin{equation}
ds^2 = a^2(\tau) [-d\tau^2 + (\delta_{ij} + h_{ij})dx^i dx^j],
\end{equation}
\begin{equation}
h_{ij} = \int d^3 k e^{i \vec{k}\cdot\vec{x}}\left(\hat{k}_i \hat{k}_j h(\vec{k},\tau)
+(\hat{k}_i \hat{k}_j - \frac{1}{3} \delta_{ij}) 6 \eta (\vec{k},\tau) \right).
\end{equation}

The perturbation is then expanded in Legendre polynomials as
\begin{equation}
\Psi = \sum_{l=0}^{\infty}(-i)^l(2l+1)\Psi_l P_l(\mu).
\end{equation}
Following \cite{Ma} we can now write the Boltzmann equation as a moment hierarchy for the $\Psi_l$
by performing the angular integration of $L[f]$
\begin{eqnarray}
\dot\Psi_0 & = & -k \frac{q}{\epsilon} \Psi_1 + \frac{1}{6} \dot{h} \frac{d \ln f_0}
{d \ln q} \label{eq:psi0}\\
\dot\Psi_1 & = & k \frac{q}{3 \epsilon}(\Psi_0 - 2 \Psi_2) \label{eq:psi1}\\
\dot\Psi_2 & = & k \frac{q}{5 \epsilon}(2 \Psi_1 - 3 \Psi_3) - \left(\frac{1}{15}
\dot{h}+\frac{2}{5}\dot\eta\right)\frac{d \ln f_0}{d \ln q} \\
\dot\Psi_l & = & k \frac{q}{(2l+1)\epsilon}(l \Psi_{l-1} - (l+1)\Psi_{l+1})
\,\,\, , \,\,\, l \geq 3
\end{eqnarray}
It should be noted here that the first two hierarchy equations are directly
related to energy and momentum conservation respectively (see e.g. \cite{Hannestad:2000gt}).

\subsection{Truncating the moment hierarchy}

Several types of approximation schemes have been tested in which neutrinos are treated as a non-perfect fluid with an approximate value of $\Psi_3$, allowing the system of equations to be truncated at order 2 (see e.g.\ \cite{Hu:1998kj,Shoji:2010hm,Lesgourgues:2011rh,Blas:2011rf}).

Truncating the hierarchy evidently makes the numerical solution of the Boltzmann equation faster. Even more importantly it will be essential for the development of numerical Boltzmann solvers in the non-linear regime, i.e.\ for the implementation of neutrinos in N-body simulations.

Before proceeding to describe an approximation for $\Psi_3$ we take a closer look at the Boltzmann hierarchy of equations. It is well known that in the absence of any gravitational source term the solution for massless neutrinos is proportional to the spherical Bessel functions of the first kind, $\Psi_l \propto j_l$. At high $l$ this approximately holds even in the presence of gravity and this property is used to truncate the Boltzmann hierarchy in numerical solvers like CLASS and CAMB.

For the pure Bessel function solution it is true that 
\begin{equation}
\Psi_{l+1} = \frac{2l+1}{\alpha \tau} \Psi_l - \Psi_{l-1}
\end{equation}
and 
one might hope that an almost unmodified version of this can be used to truncate the real hierarchy at $l=2$, i.e.\ that
\footnote{We note that, as discussed in detail in Ref.~\cite{Lesgourgues:2011rh}, this expression is not gauge independent, unlike the real $\Psi_3$.}.
\begin{equation}
\Psi_3  \simeq  \frac{5 q}{k \epsilon \tau} \Psi_2 - \Psi_{1, {\rm Sync.}}.
\label{eq:mbapprox}
\end{equation}

\begin{figure}[h]
\begin{tabular}{ll}
\includegraphics[width=8cm]{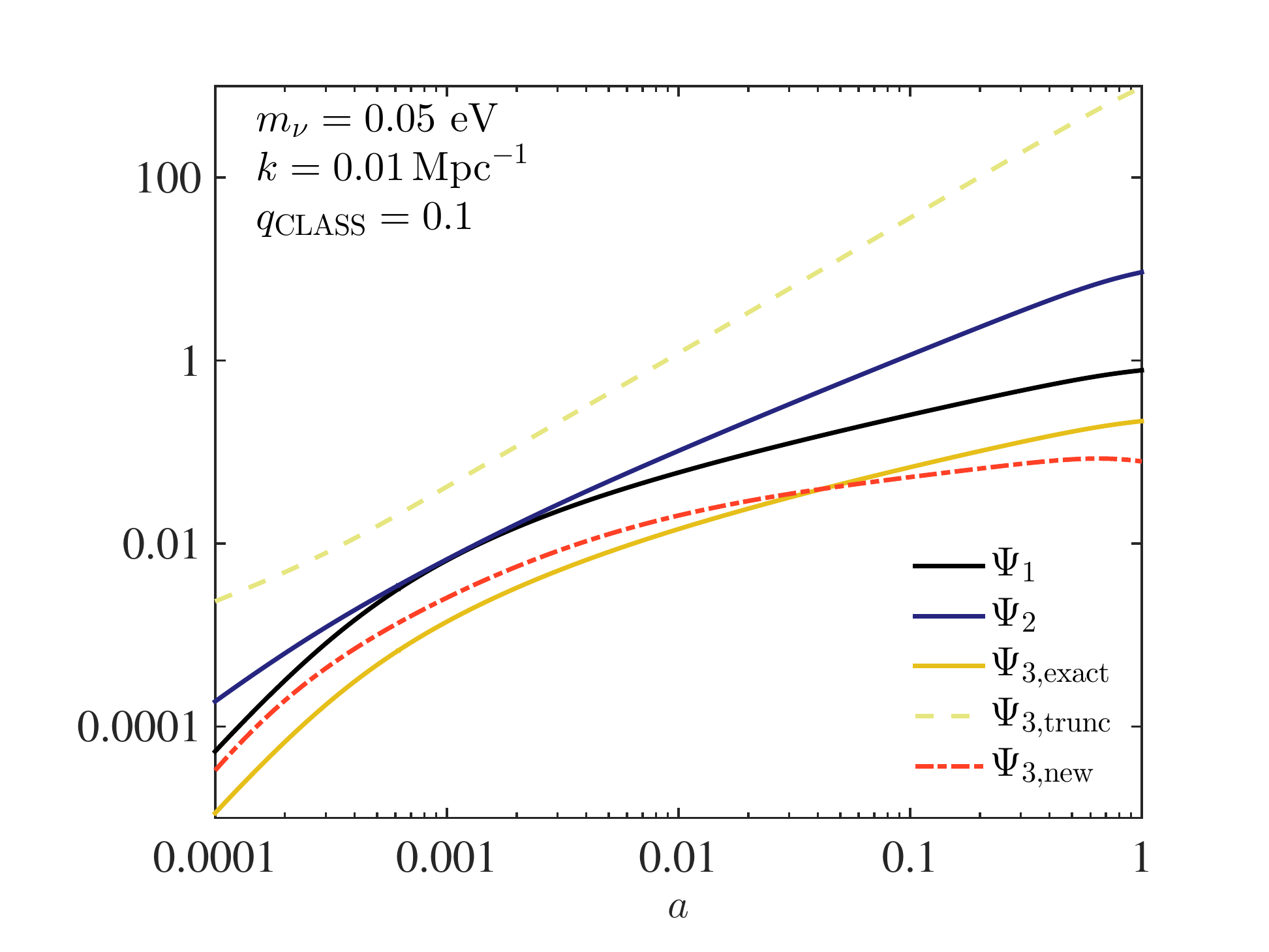}&\includegraphics[width=8cm]{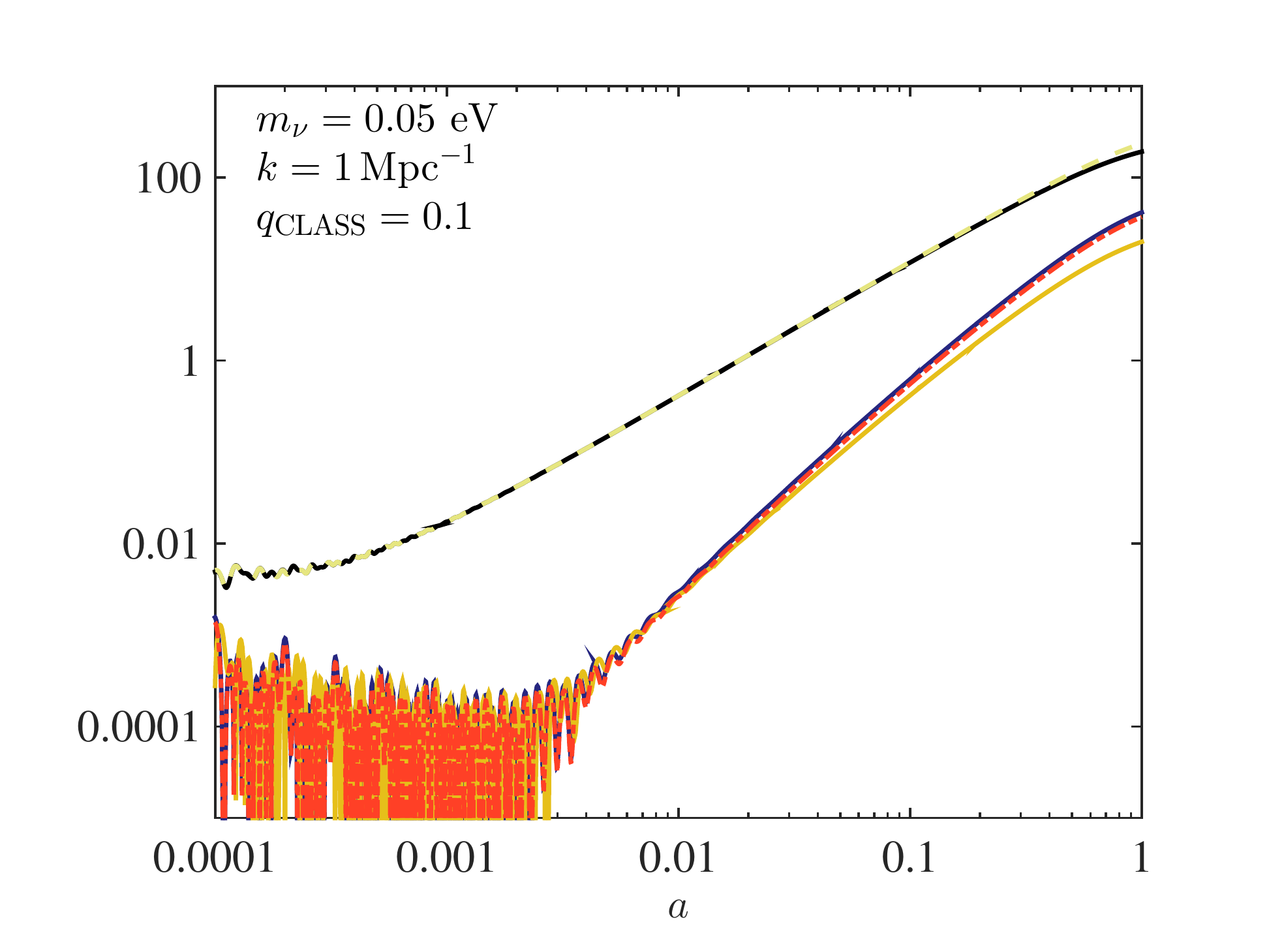}\\
\includegraphics[width=8cm]{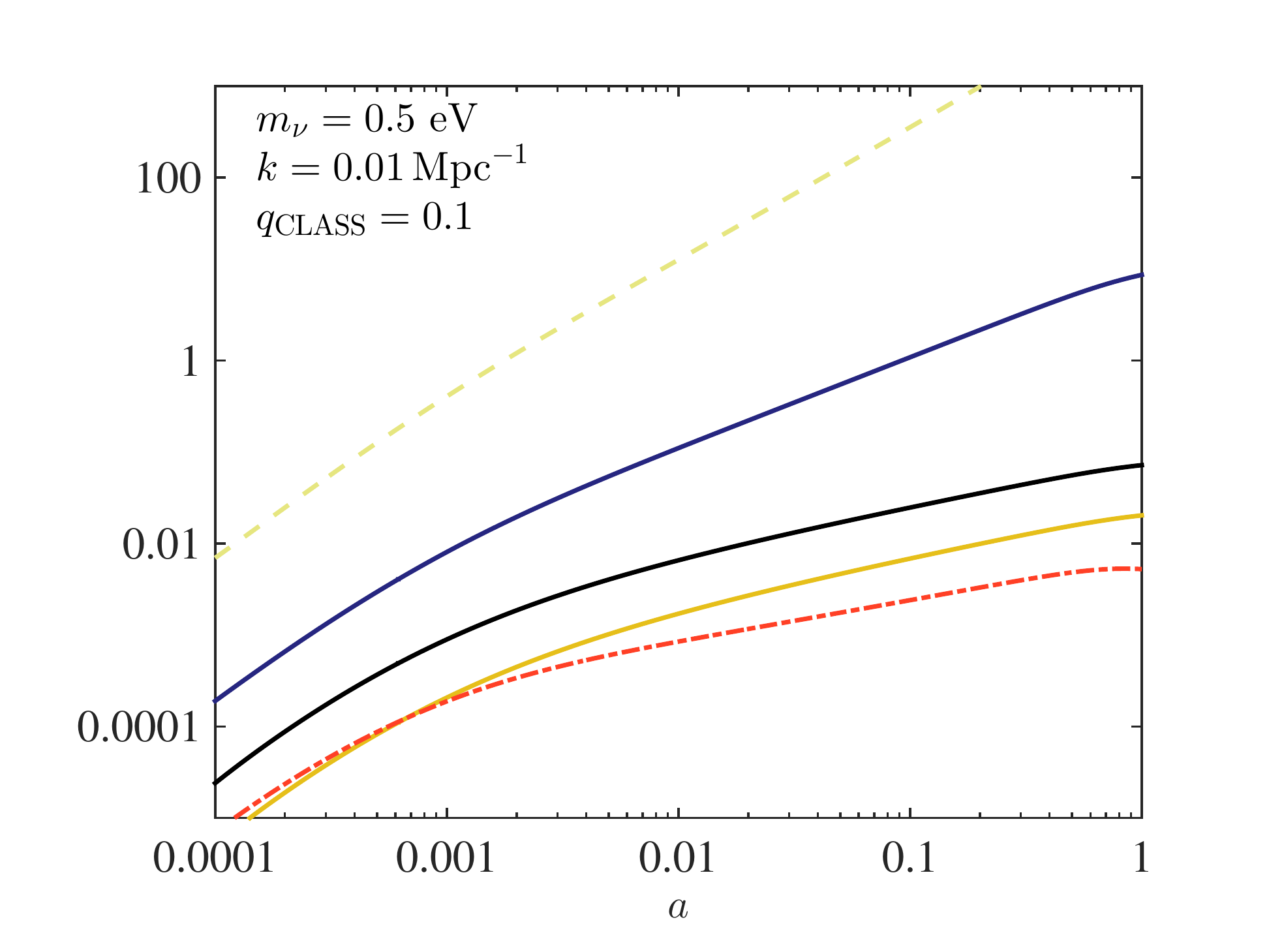}&\includegraphics[width=8cm]{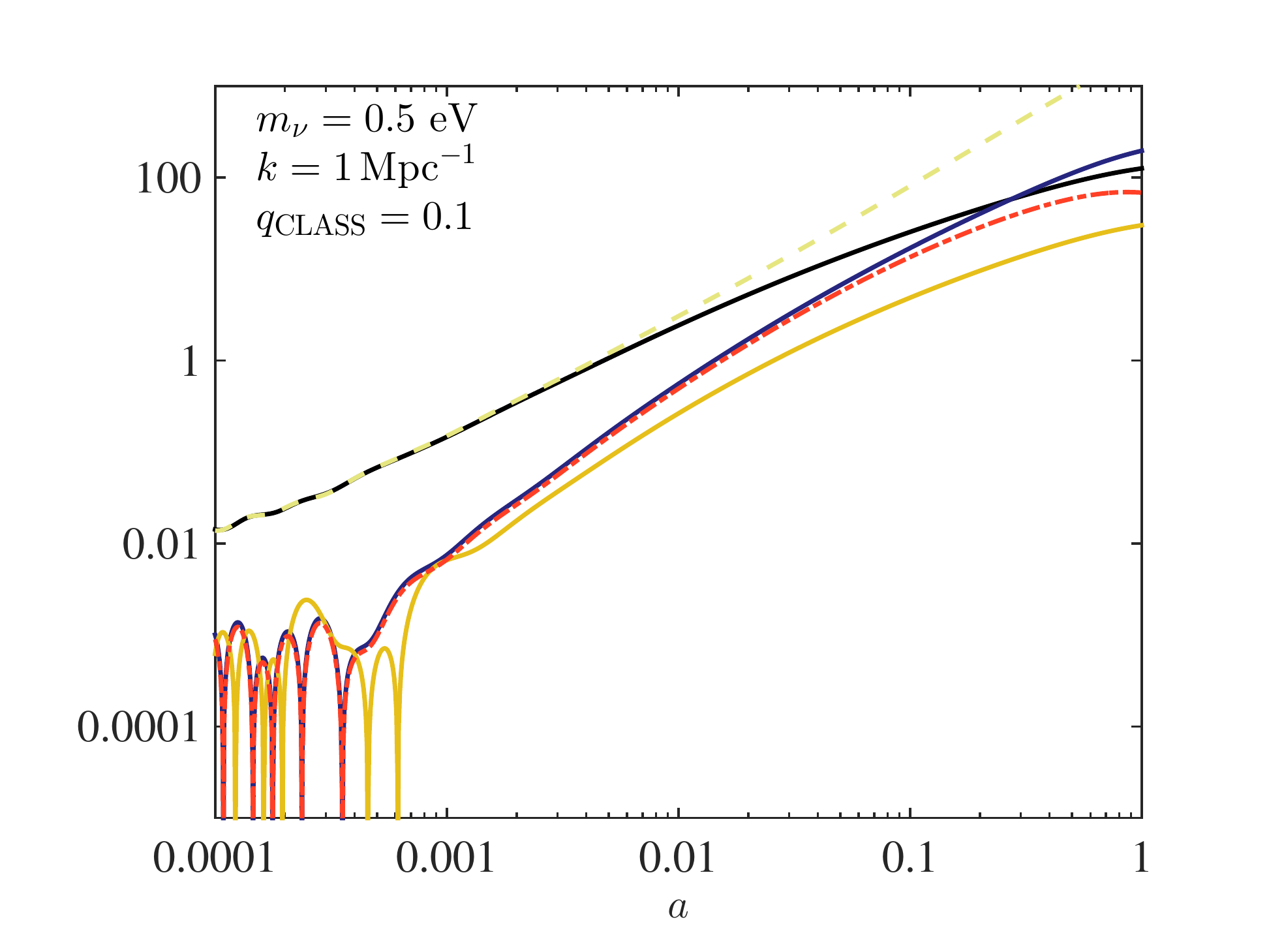}\\
\end{tabular}
\caption{$\lvert \Psi_{1, {\rm Sync}} \rvert$ (black solid lines), $\lvert \Psi_2 \rvert$ (blue solid lines), $\lvert \Psi_3 \rvert$ (orange solid lines), and the approximation from Eq.~\ref{eq:mbapprox} ($\Psi_{3, {\rm trunc}}$, yellow dashed lines) and from Eq.~\ref{eq:psi3approx} ($\Psi_{3, {\rm new}}$, red dash-dot lines) for $m_\nu = 0.5$ eV, $k=0.01 \, \textrm{Mpc}^{-1}$ (top left panel),  $m_\nu = 0.5$ eV, $k=1 \, \textrm{Mpc}^{-1}$ (top right panel), $m_\nu = 0.05$ eV, $k=0.01 \, \textrm{Mpc}^{-1}$ (bottom left panel), $m_\nu = 0.05$ eV, $k=1 \, \textrm{Mpc}^{-1}$ (bottom right panel).
The momentum bin is $q_{\rm CLASS}=0.1$.
\label{fig:psis}}
\end{figure}

\begin{figure}[h]
\begin{tabular}{ll}
\includegraphics[width=8cm]{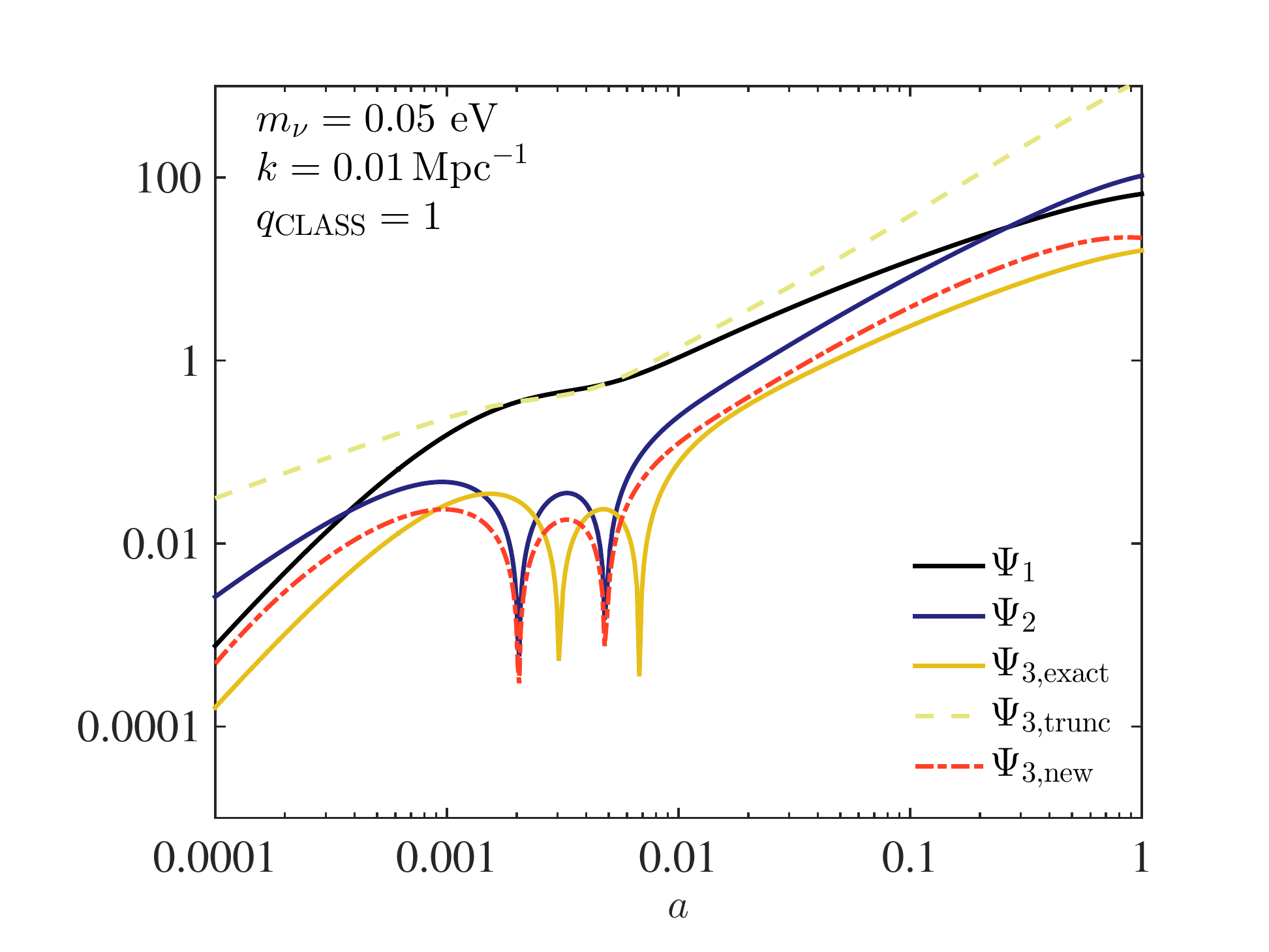}&\includegraphics[width=8cm]{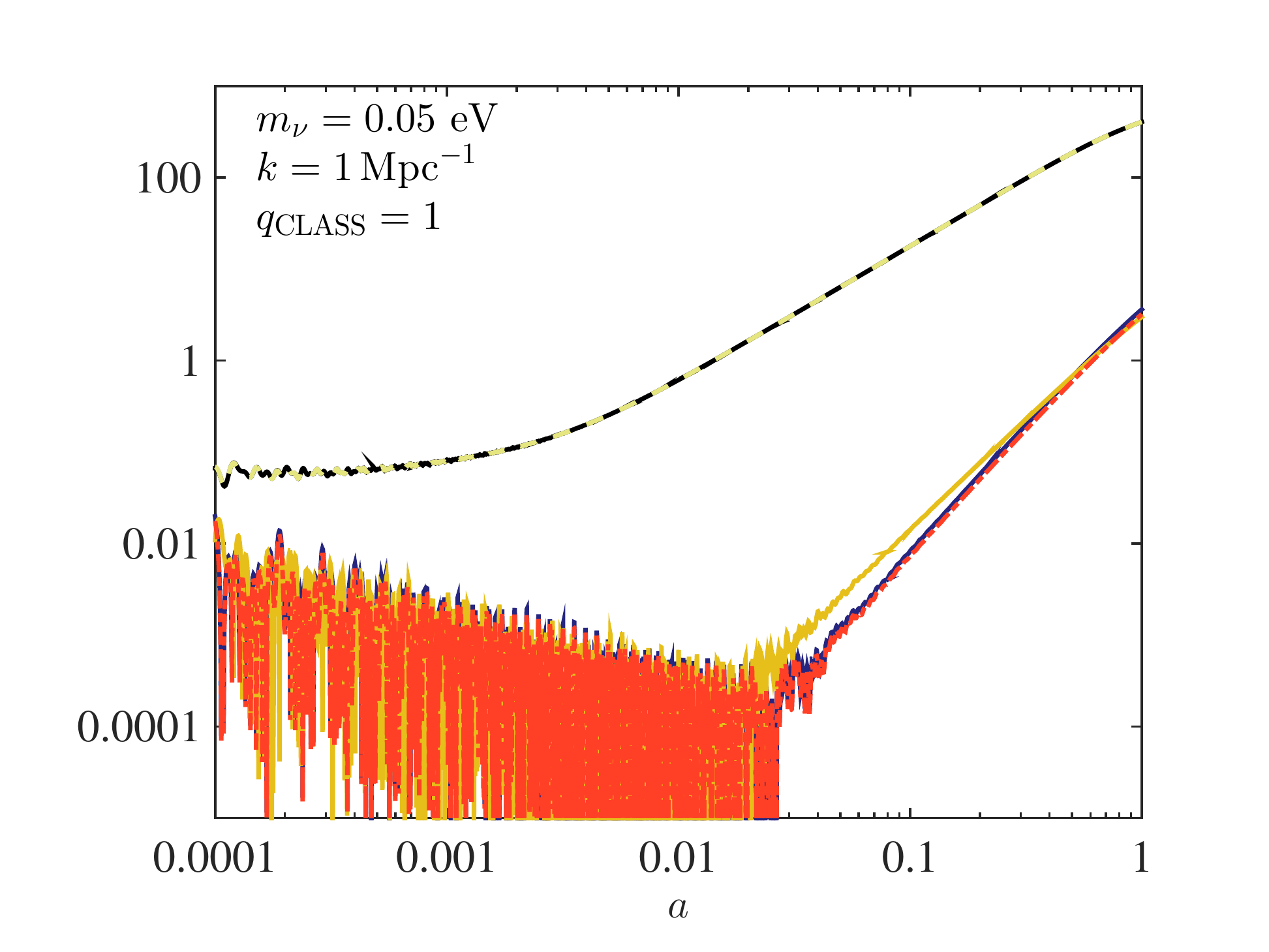}\\
\includegraphics[width=8cm]{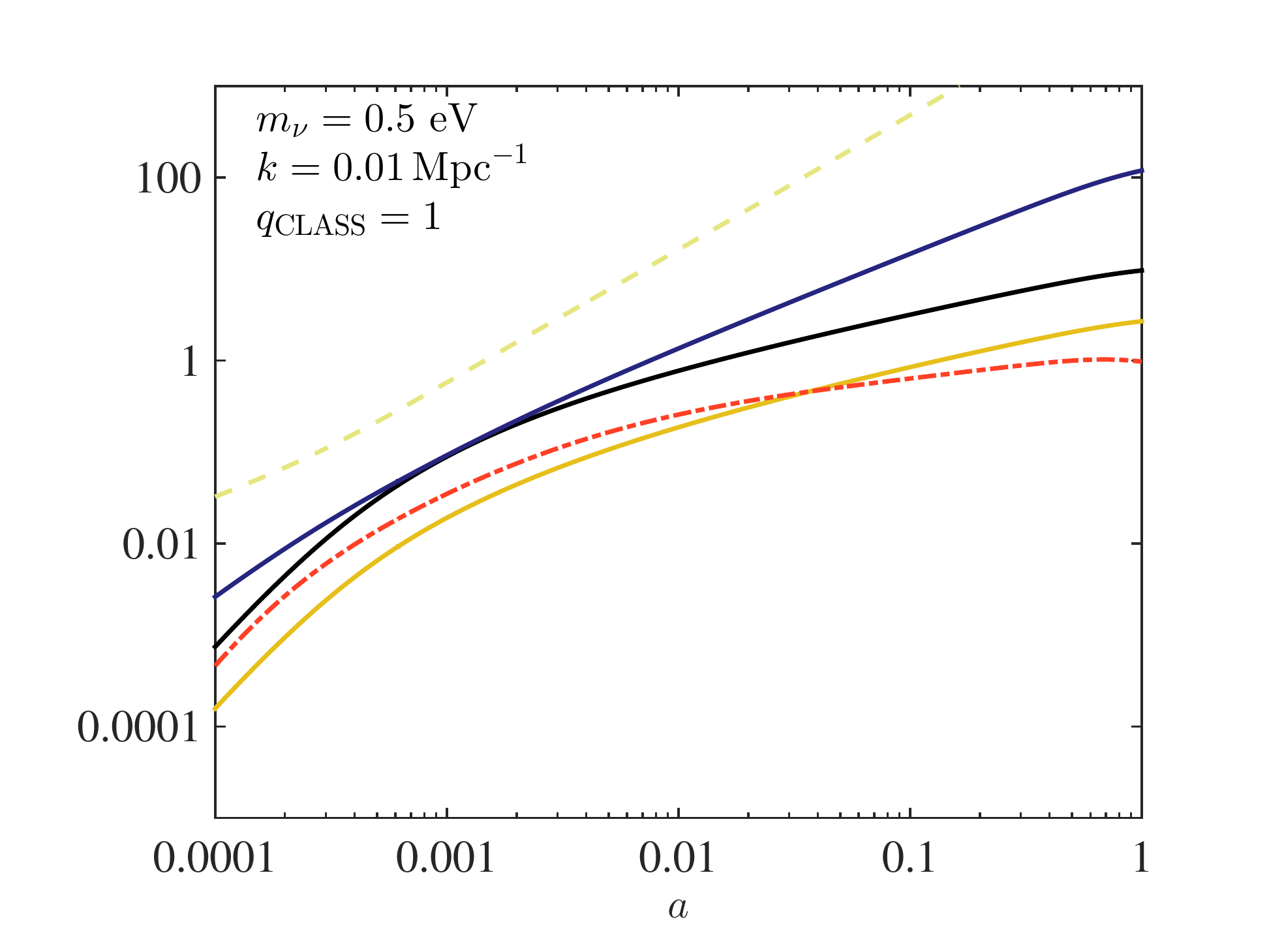}&\includegraphics[width=8cm]{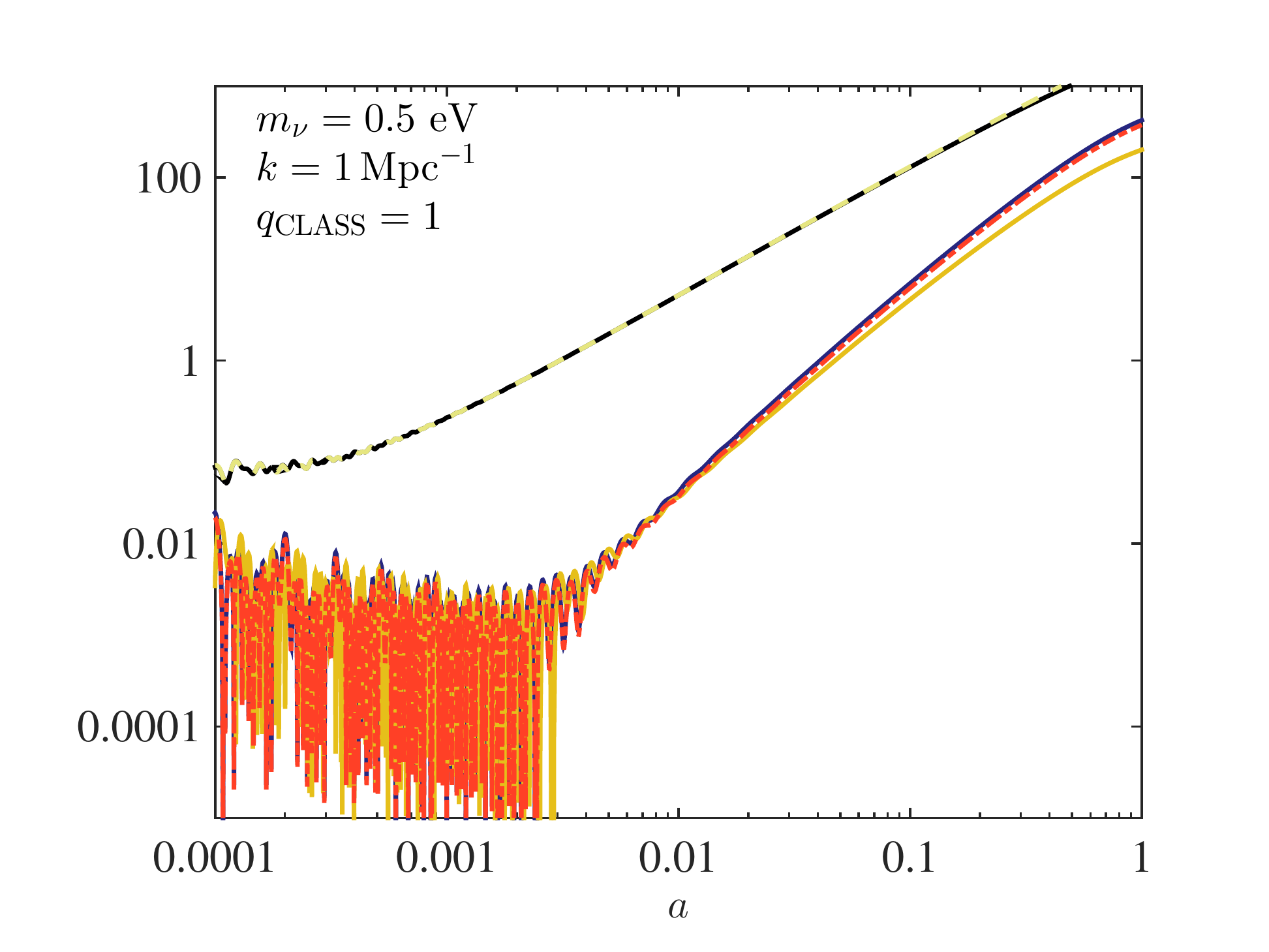}\\
\end{tabular}
\caption{Same as in Fig.~\ref{fig:psis} but for a different momentum bin: $q_{\rm CLASS}=1$, i.e. $q \sim T_{\nu}$.\label{fig:psis2}}
\end{figure}

However, because of the gravitational source terms we cannot a priori expect this approximation to be accurate. To illustrate the potential problem with this we plot $\Psi_1, \Psi_2, \Psi_3$, as well as the approximation above (labeled as $\Psi_{3,{\rm trunc}}$) as functions of $a$ for two different values of $q$, $k$ and $m_\nu$ in Figs.~\ref{fig:psis} and \ref{fig:psis2}
\footnote{This, and all other numerical calculations presented in this paper were done with the CLASS code, but we have tested that results are identical if calculated using CAMB.}.
As can be seen this approximation is generally very poor.

Instead we can let ourselves be guided by the same figure towards a much better approximation for $\Psi_3$ by noting that for most values of $k$ and $m_\nu$ the relation $\Psi_3 \sim \Psi_2$ seems to hold remarkably well.

This is easy to understand from the following analytic considerations. In the absence of gravity and in a non-expanding universe the
Boltzmann equation hierarchy can be simply written as
\begin{equation}
\dot \Psi_l = \frac{\alpha}{2l+1} [l \Psi_{l-1} - (l+1) \Psi_{l+1}].
\end{equation}
The solution to this system of equations is $\Psi_l \propto j_l(\alpha \tau)$, i.e.\ a set of functions oscillating with the period $1/\alpha$ and damped as $1/(\alpha \tau)$ at large $\tau$. From this we get that the envelope of $\Psi_3$ is identical to that of $\Psi_2$, i.e. at large $\tau$ they differ only by a phase in the absence of gravity.

When the gravitational source terms are important the solutions are still quite similar in nature, and in order to better appreciate
the nature of the solutions we study a very simple toy model before returning to the real system of equations.

More specifically we look at the system of equations given by
\begin{equation}
\dot \Phi_l = \frac{\alpha}{2l+1} [l \Phi_{l-1} - (l+1) \Phi_{l+1}] + (\delta_{l0}+\delta_{l2}) f(\tau),
\label{eq:test}
\end{equation}
where $f(\tau)$ is an $l$-independent function. This resembles (but is not equal to) the true Boltzmann hierarchy. We will also assume that $\alpha$ is time-independent which is not generally true for the true Boltzmann hierarchy.

At late times $(\alpha \tau \gg 1)$ the solution to the toy model Boltzmann hierarchy for $l > 1$ is
\begin{equation}
\Phi_l \propto g(\tau)/\sqrt{2l+1},
\label{eq:solution1}
\end{equation}
where $g(\tau)$ is a common $l$-independent function. Since we are aiming at an expression for $\Phi_3$ in terms of lower moments the important point to take away from Eq.~(\ref{eq:solution1}) is that for $(\alpha \tau \gg 1)$ 
\begin{equation}
\Phi_3 \sim \sqrt{\frac{5}{7}} \Phi_2.
\label{eq:largex}
\end{equation}

Conversely, when the argument is small $(\alpha \tau \alt 1)$ the solution to Eq.~(\ref{eq:test}) is such that 
\begin{equation}
\Phi_l \sim \frac{\alpha \tau}{2 l + 1} \Phi_{l-1}, 
\end{equation}
which implies that
\begin{equation}
\Phi_3 \sim \frac{\alpha \tau}{7} \Phi_{2}. 
\label{eq:smallx}
\end{equation}

\begin{figure}[h]
\begin{center}
\includegraphics[width=10cm]{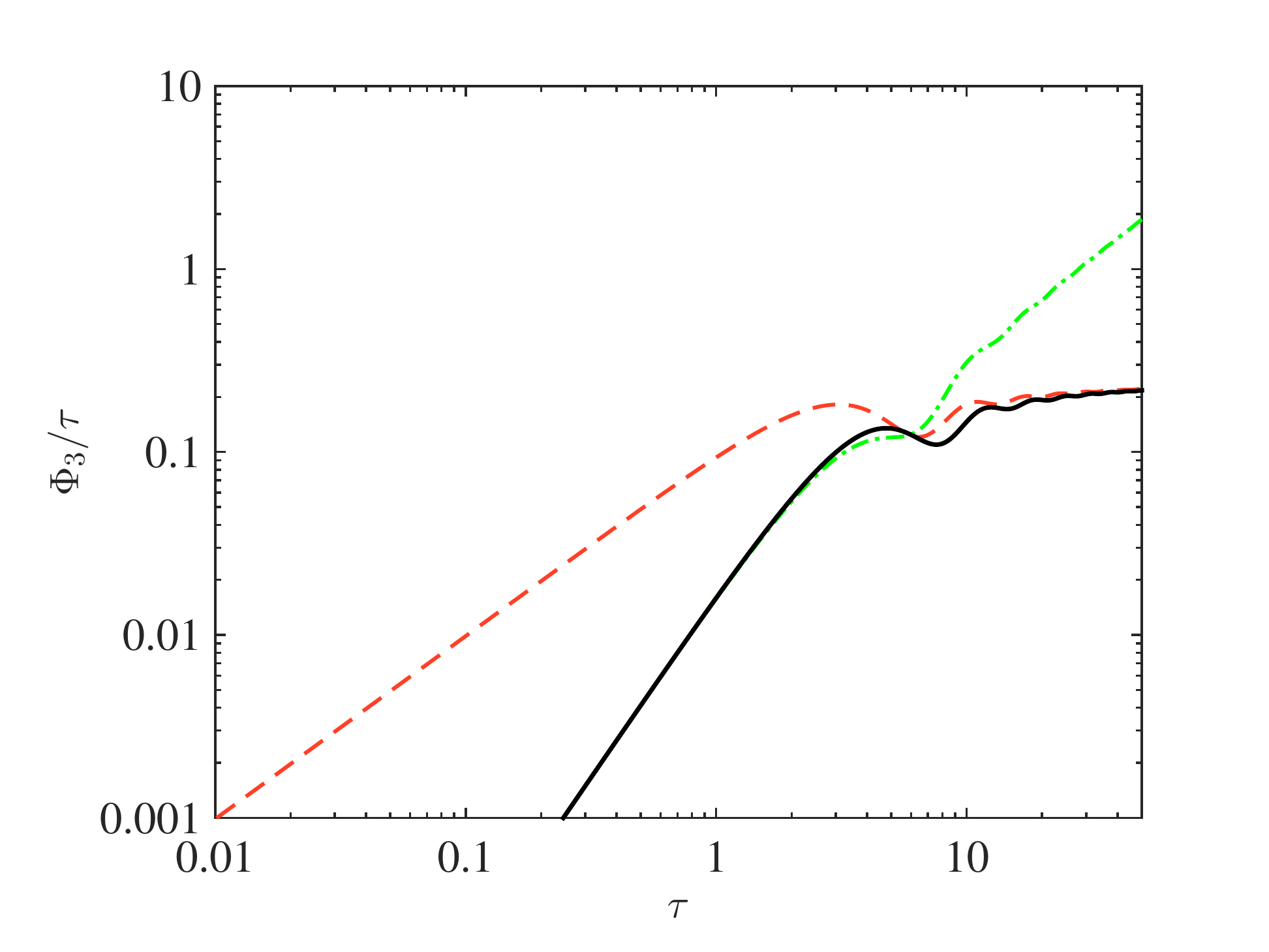}
\end{center}
\caption{The solution to $\Phi_3$ from the toy model in Eq.~(\ref{eq:test2}) (black line). We also show the asymptotic solutions from
Eq.~(\ref{eq:smallx}) (green dash-dot line) and Eq.~(\ref{eq:largex}) (red dashed line).
\label{fig:bessel}}
\end{figure}

We show a simple example of this in Fig.~\ref{fig:bessel} for the case where $\alpha=1$ and where the driving term is such that
\begin{equation}
\dot \Phi_l = \frac{\alpha}{2l+1} [l \Phi_{l-1} - (l+1) \Phi_{l+1}] + 0.1 \, (\delta_{l0}+\delta_{l2}) \tau,
\label{eq:test2}
\end{equation}
and the initial condition is given by $\Phi_0(0)=1$, $\Phi_{l \neq 0} = 0$.

In the same figure we also show the approximate solutions given in Eqs.~(\ref{eq:largex}) and (\ref{eq:smallx}). In the specific case shown the transition between the two asymptotic solutions occurs at $\alpha \tau \sim$ few.

Based on the solutions to the toy model hierarchy we might now guess at a solution to the true hierarchy of the form

\begin{equation}
\Psi_3 \simeq  \left( \frac{\frac{x}{7} \frac{\beta}{x} +  \sqrt{\frac{5}{7}} \frac{x}{\beta}}{\frac{\beta}{x}+\frac{x}{\beta}} \right) \, \Psi_2,
\end{equation}
with $x \equiv k q \tau/\epsilon$, and where $\beta$ is a numerical constant controlling the point of transition between the two asymptotic solutions.

Numerically we find that this approximation works extremely well with $\beta=1$, and that if a small $k$-dependent correction is included it works even better
\begin{equation}
\Psi_3 \simeq \left( \frac{\frac{1}{7}+ \sqrt{\frac{5}{7}} x}{\frac{1}{x}+x} \right) \,  \left(\frac{k}{1 \, h/{\rm Mpc}}\right)^{0.12} \, \Psi_2.
\label{eq:psi3approx}
\end{equation}
In Figs.~\ref{fig:psis} and \ref{fig:psis2} we show this approximation (labeled as $\Psi_{3,{\rm new}}$) as a function of $a$ for two different values of $q$, $k$ and $m_\nu$.

\begin{figure}[h]
\begin{tabular}{lll}
\includegraphics[width=5cm]{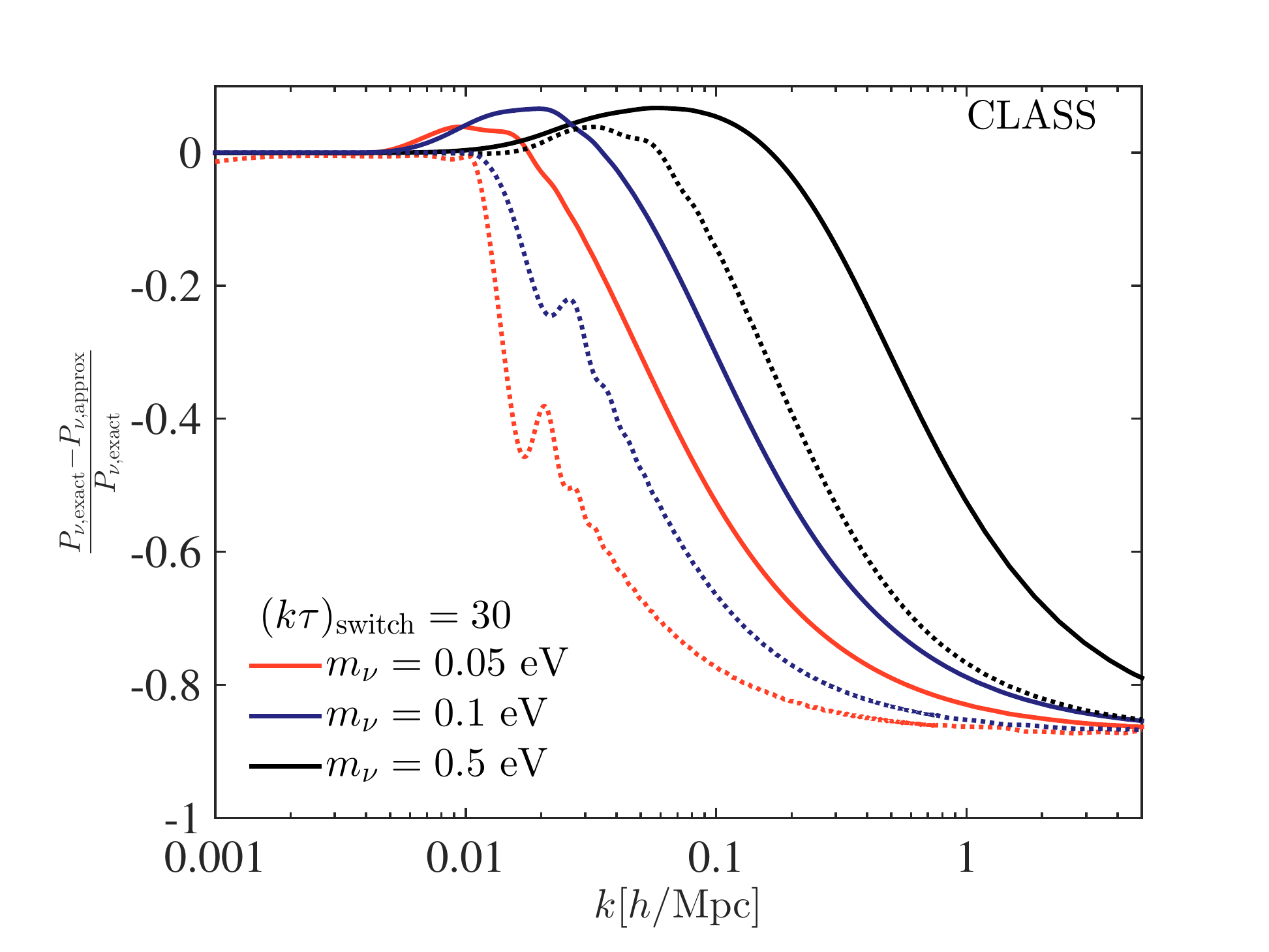} & \includegraphics[width=5cm]{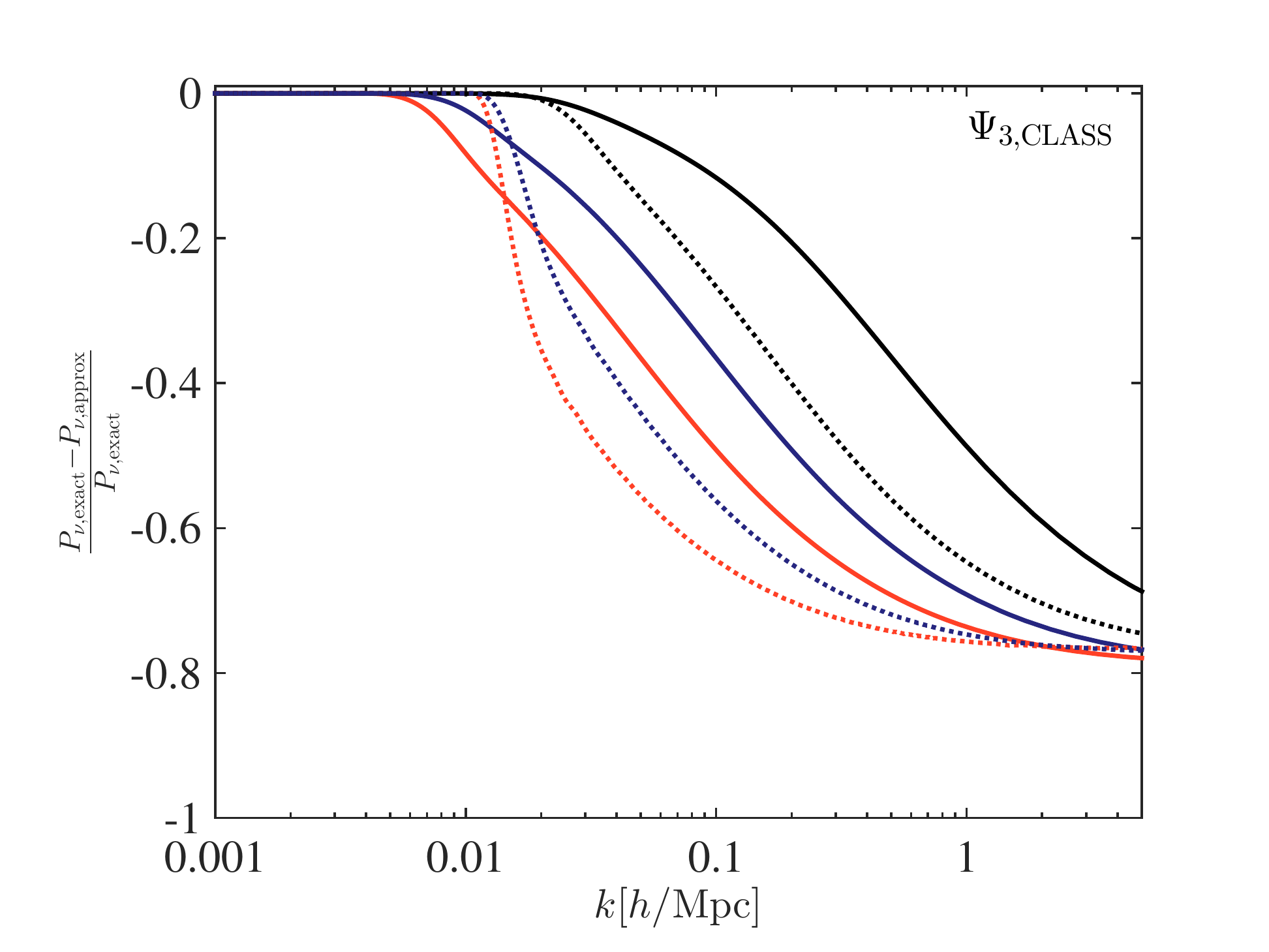}
&\includegraphics[width=5cm]{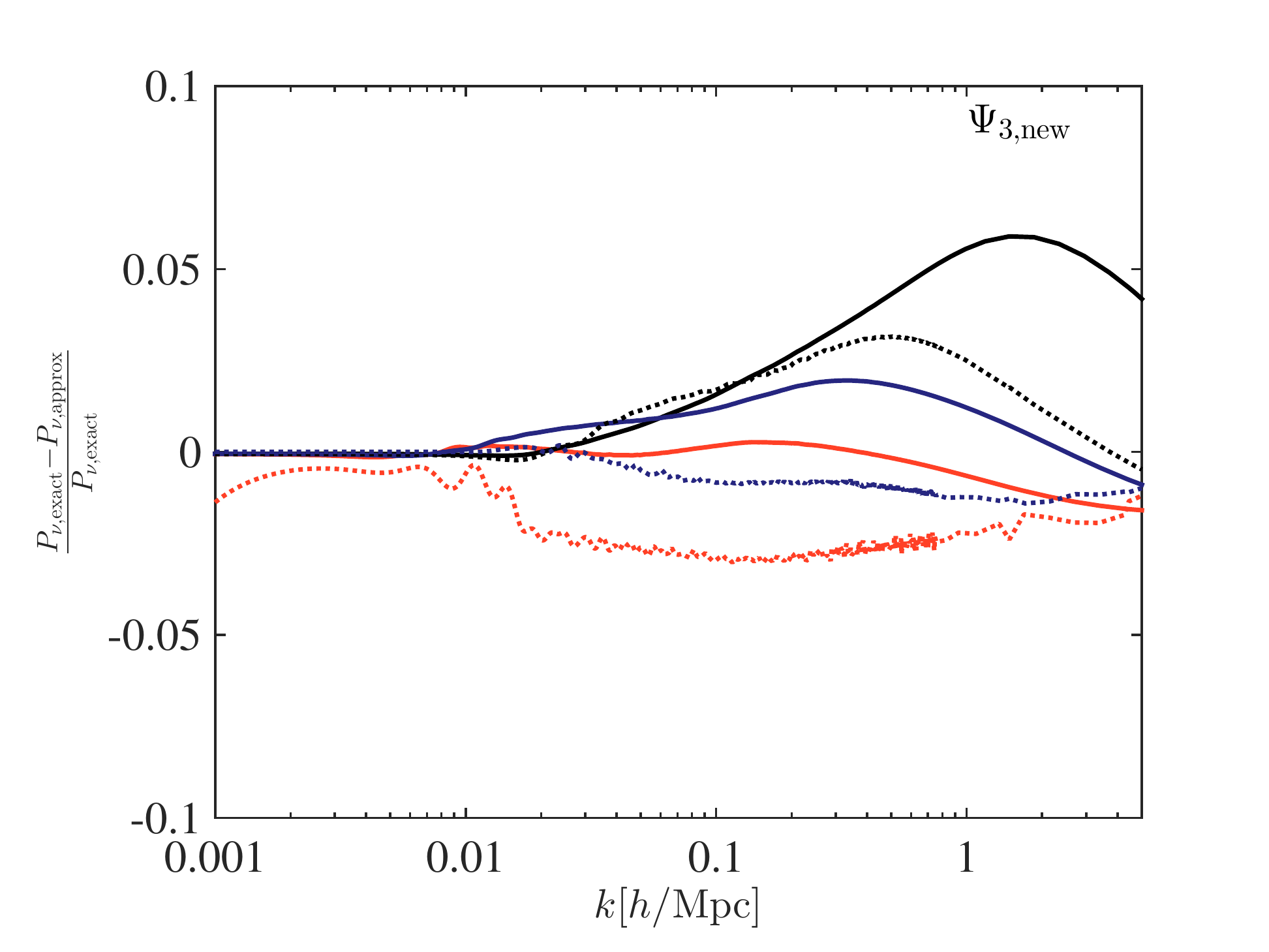} \\ 
\includegraphics[width=5cm]{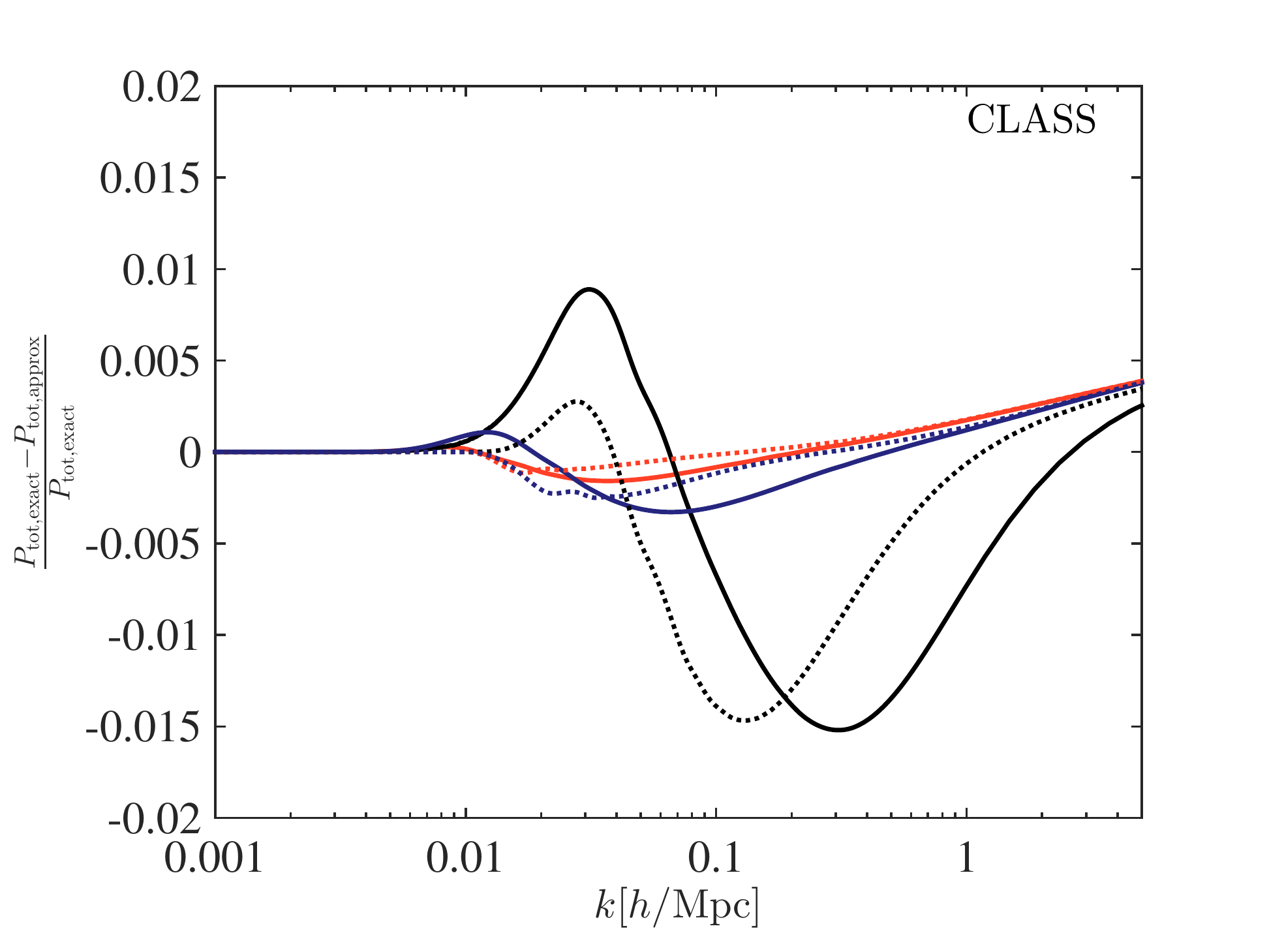} & \includegraphics[width=5cm]{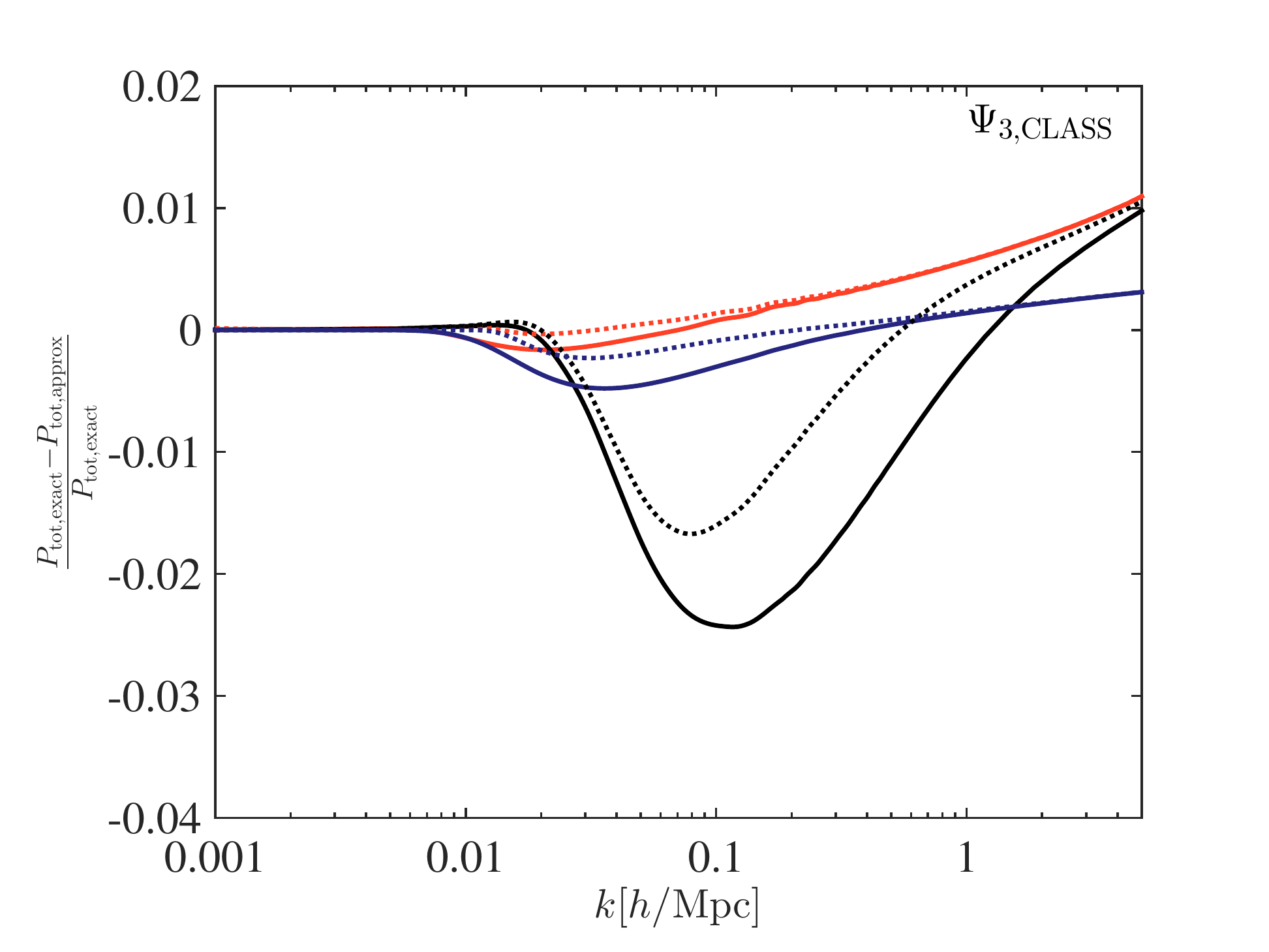}
&\includegraphics[width=5cm]{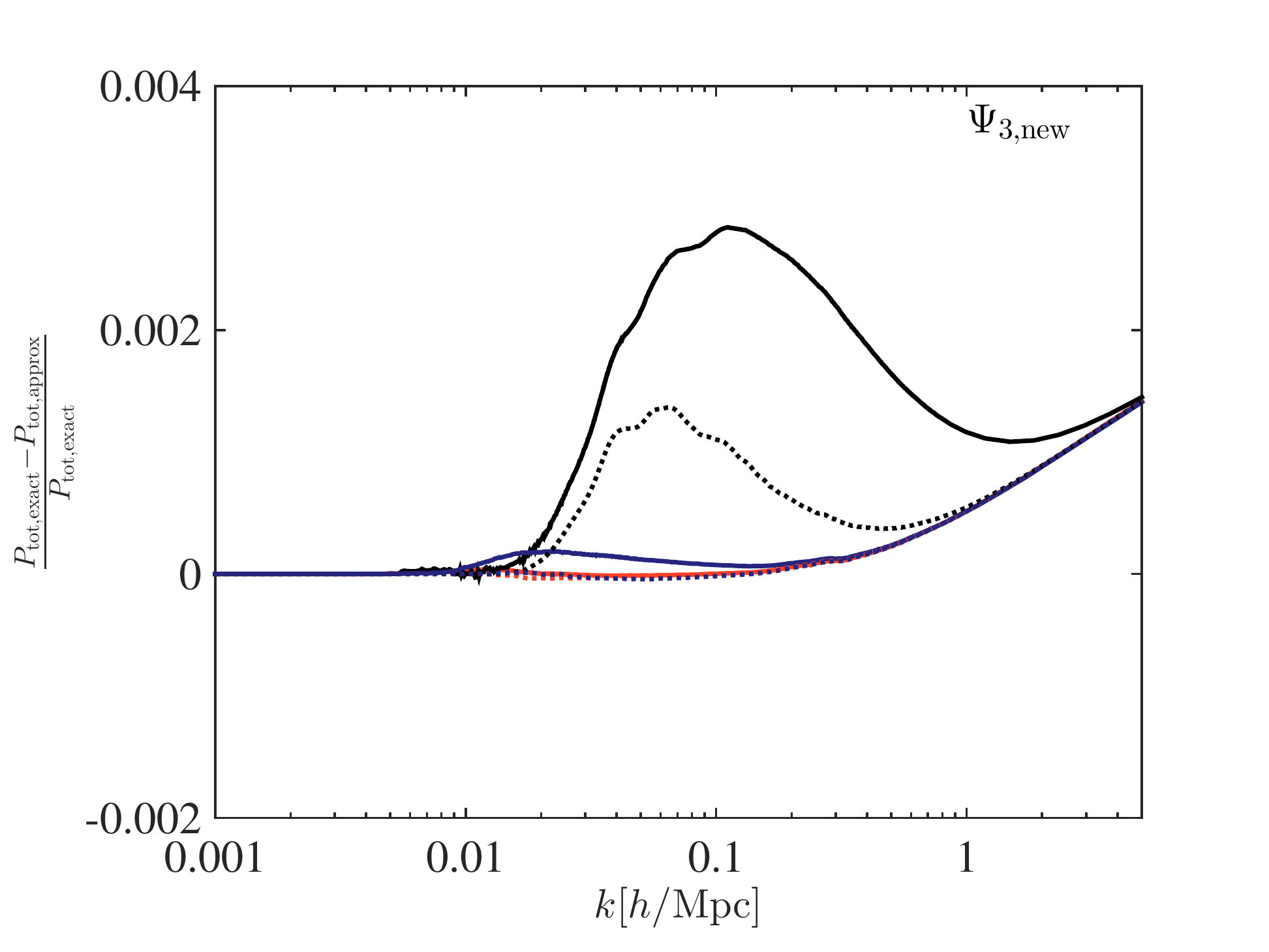} \\ 
\includegraphics[width=5cm]{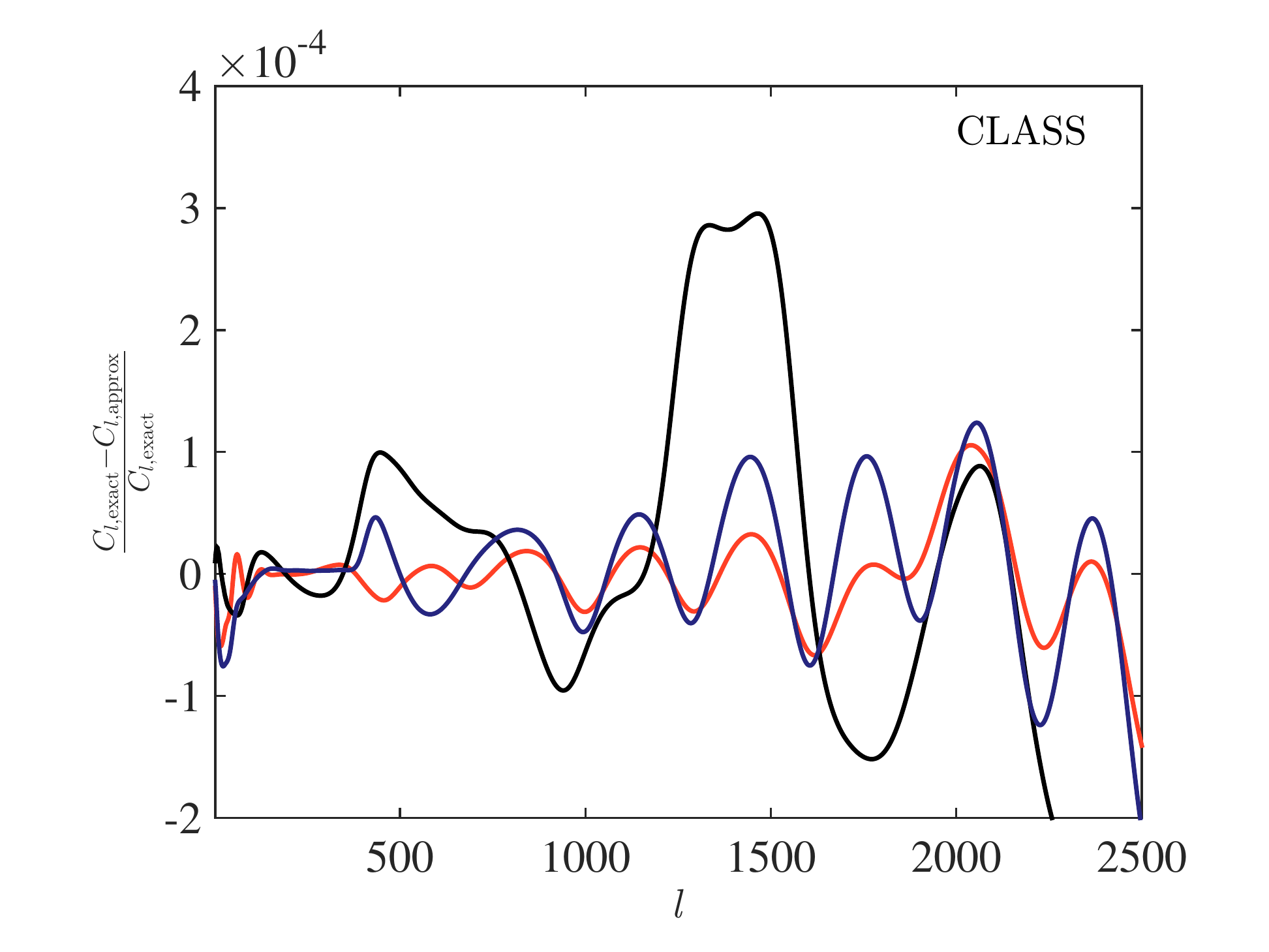} & \includegraphics[width=5cm]{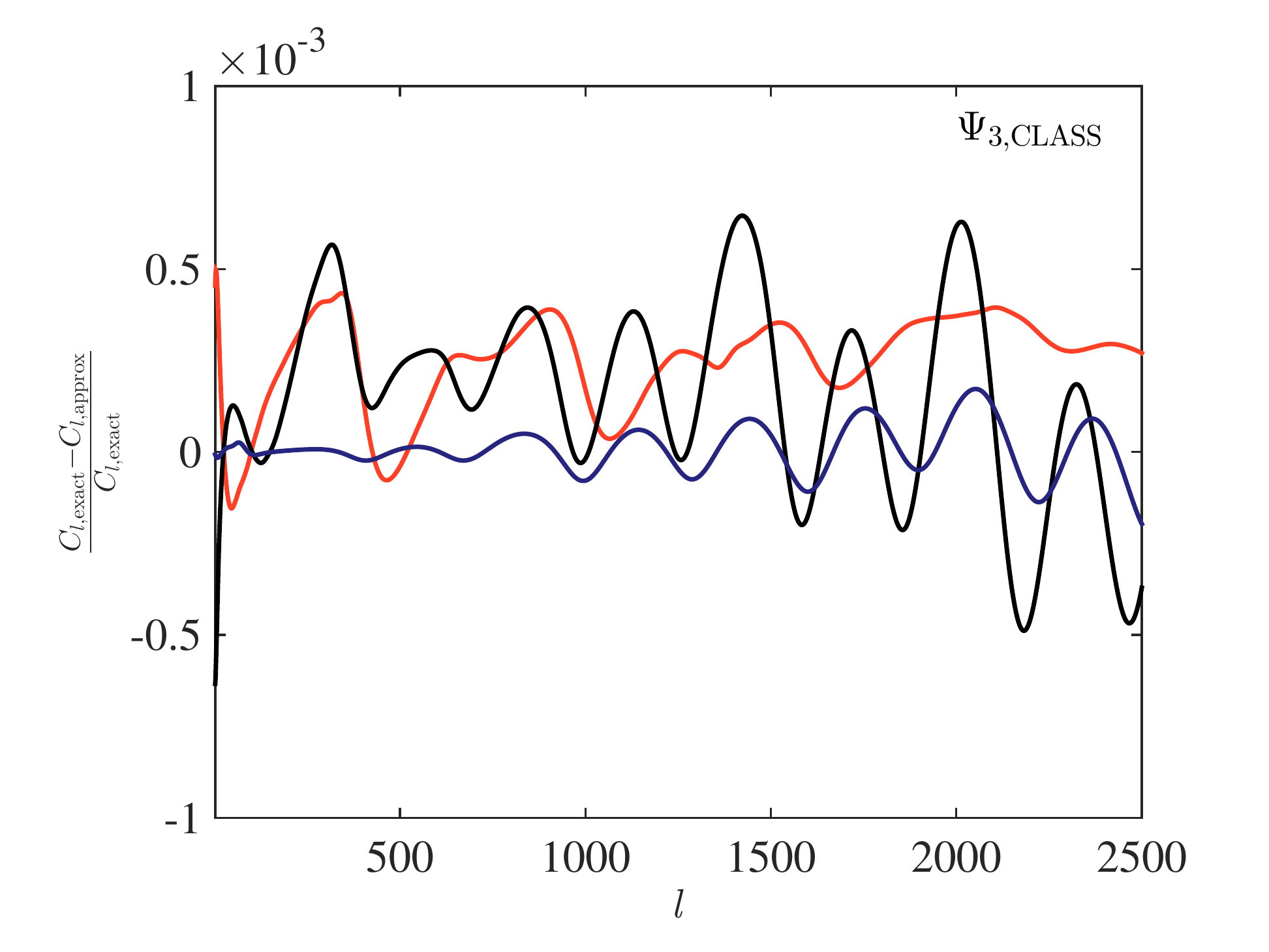}
&\includegraphics[width=5cm]{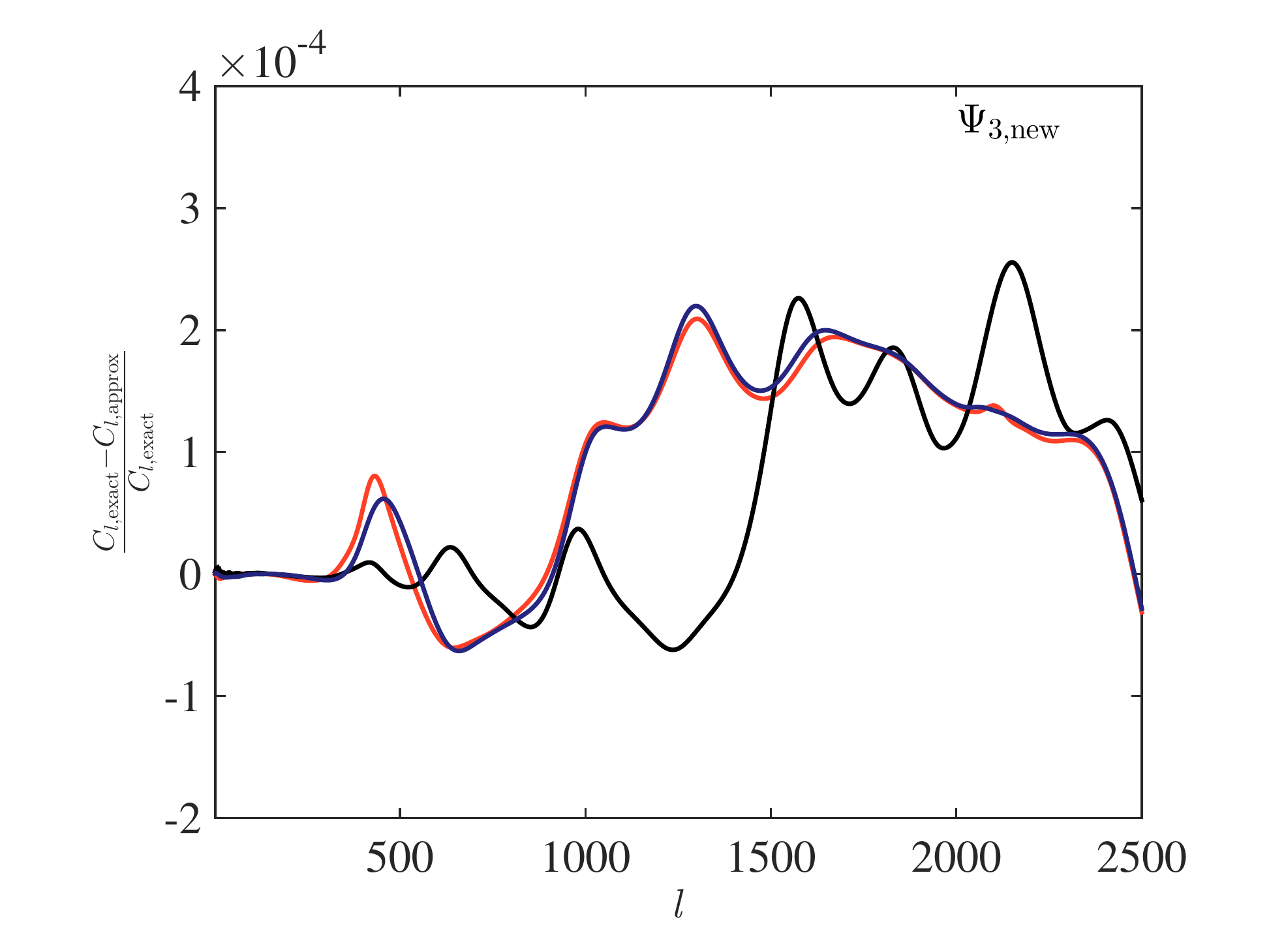} \\ 
\end{tabular}
\caption{The ratio $(P_{\nu,{\rm approx}}-P_{\nu, {\rm exact}})/P_{\nu, {\rm exact}}$ (upper panels)
and the ratio $(P_{\rm tot, approx}-P_{\rm tot, exact})/P_{\rm tot, exact}$ (middle panels)
and the ratio $(C_{\ell,{\rm approx}}-C_{\ell, {\rm exact}})/C_{\ell, {\rm exact}}$ (lower panels)
using CLASS fluid approximation (left panels), truncating the hierarchy at $\ell=2$ with the CLASS version of $\Psi_3$ (middle panels) and with Eq.{eq:psi3approx} (left panels).
In all the cases the switch of the approximation/truncation is at $k\tau>30$.
The ratios are given for various neutrino masses $m_\nu=0.5$ eV (black lines), 0.1 eV (blue lines), and 0.05 eV (red lines) and at different redshifts $z=10$ (dotted lines) and $z=0$ (full lines). 
\label{fig:class_diff}}
\end{figure}

\subsection{Results}

In the right panel of Fig.~\ref{fig:class_diff} we show the difference using the approximation in Eq.~(\ref{eq:psi3approx}) and the result obtained from the full Boltzmann hierarchy. In practise the full hierarchy is solved until $k \tau = 30$ at which point the code switches to using the approximation. From the top panel we see that the neutrino power spectrum itself can be calculated at better than $\sim$ 5\% precision for all the masses and redshifts investigated. The total matter power spectrum has an error of less than 0.5\% - more than adequate for the analysis of future high-precision data.
Finally, the CMB spectrum picks up only minute errors, far smaller than what can be detected even with cosmic variance limited experiments. The main reason is that the switch from solving the full hierarchy to using the approximation happens at $k \tau=30$. If the switch is made earlier the error increases.

For comparison we also show two other cases: The left column is obtained with the standard ``CLASS'' approximation
\cite{Lesgourgues:2011rh} in which the hierarchy is truncated at $l=2$ {\it and} integrated over momentum space. The middle column shows results for the case where the ``CLASS'' approximation for $\Psi_3$ is used, but the momentum dependence is kept in the Boltzmann equation.

\begin{figure}[h]
\begin{center}
\includegraphics[width=10cm]{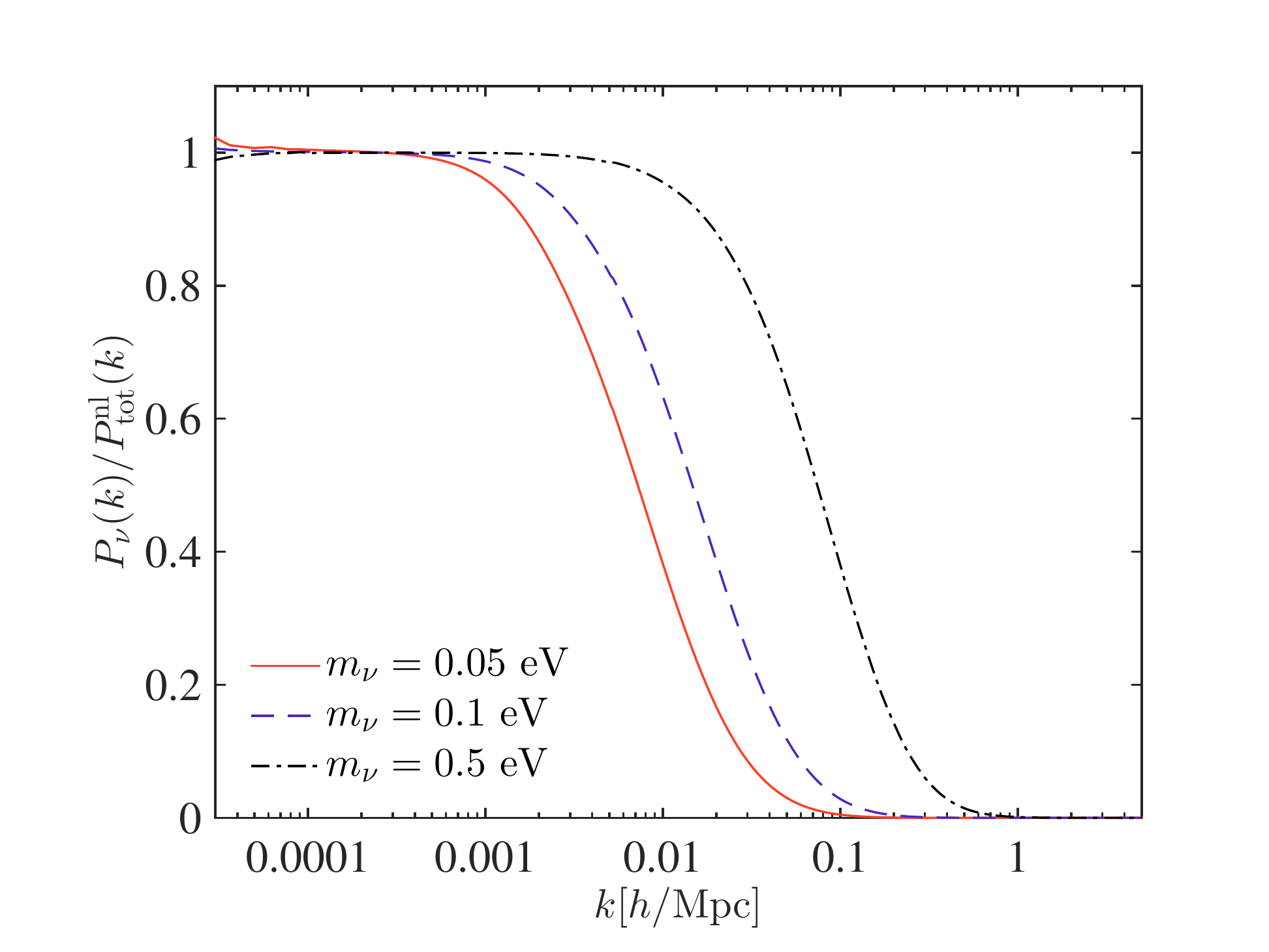}
\end{center}
\caption{Ratio between the neutrino component and the total matter power spectrum for vaious neutrino masses at $z=0$. \label{fig:ratio}}
\end{figure}

Clearly, the ``CLASS'' approximation for $\Psi_3$ overestimates $\Psi_3$ because of the $\Psi_1$ term, leading to a substantial suppression of power at high $k$. In fact the error is close to a factor 5 for both the integrated and the momentum dependent cases.
However, the total matter power spectrum is predicted with a much higher precision. The reason is easy to understand from Fig.~\ref{fig:ratio}. Even for relatively high neutrino masses the contribution to the total power spectrum of neutrinos at high $k$ is essentially null. Even if the neutrino power spectrum is wrong by a factor of five it has only a modest impact on the total power spectrum.
Still, the approximation used here fares far better than the one used by CLASS, leading to errors a factor of typically 5 smaller.

Finally, the error on the CMB spectrum is comparable between all three approximations because the switch is triggered at high $k \tau$ where most of the CMB physics has already happened. We also note that the errors on the CMB spectra from using the approximation with this swtich are substantially smaller than the differences between the CAMB and CLASS codes \cite{Lesgourgues:2011rg} and in any case completely negligible.

In principle further approximations can be made by integrating the Boltzmann hierarchy over momentum space to yield the fluid equations
(see e.g.\ \cite{Hu:1998kj,Shoji:2010hm,Lesgourgues:2011rh}). These can then be truncated by use of an approximation similar to Eq.~\ref{eq:psi3approx} for $\Psi_3$. However, in order to work with the integrated quantities several other approximations are necessary and
in this case the precision with which the neutrino power spectrum is calculated degrades very significantly
\cite{Hu:1998kj,Shoji:2010hm,Lesgourgues:2011rh}. The matter and CMB spectra in many cases can still be calculated with adequate accuracy simply because light neutrinos contribute very little to these quantities.

Keeping the momentum dependent equations rather than integrating them to get the corresponding fluid equations of course slows down the speed of computations. 
We find that running CLASS with a single thread (i.e.\ without OpenMP) the runtime when using our approximation and switching at $k\tau = 30$ is 0.43 times the runtime of CLASS when solving the full hierarchy (having $\ell_{\rm MAX}=17$ and 5 momentum bins). Using the momentum integrated fluid equations with the same switch in $k \tau$ produces a runtime of 0.22 times the runtime with the full hierarchy.
Our approximation thus leads to a speed-up of a factor 2.32 while the fluid equations yield a speed-up of a factor 4.54 \footnote{Running with multiple threads will reduce the speed-up factor because the code spends less walltime on parallel tasks.}.

\section{Entering the non-linear regime}

Having demonstrated that the neutrino matter power spectrum can be calculated to a precision of a few percent even when the Boltzmann hierarchy is truncated at $l=2$ we will proceed to discuss a simple calculation of the neutrino matter power spectrum using the full non-linear gravitational potential.

\begin{figure}[h]
\begin{center}
\includegraphics[width=10cm]{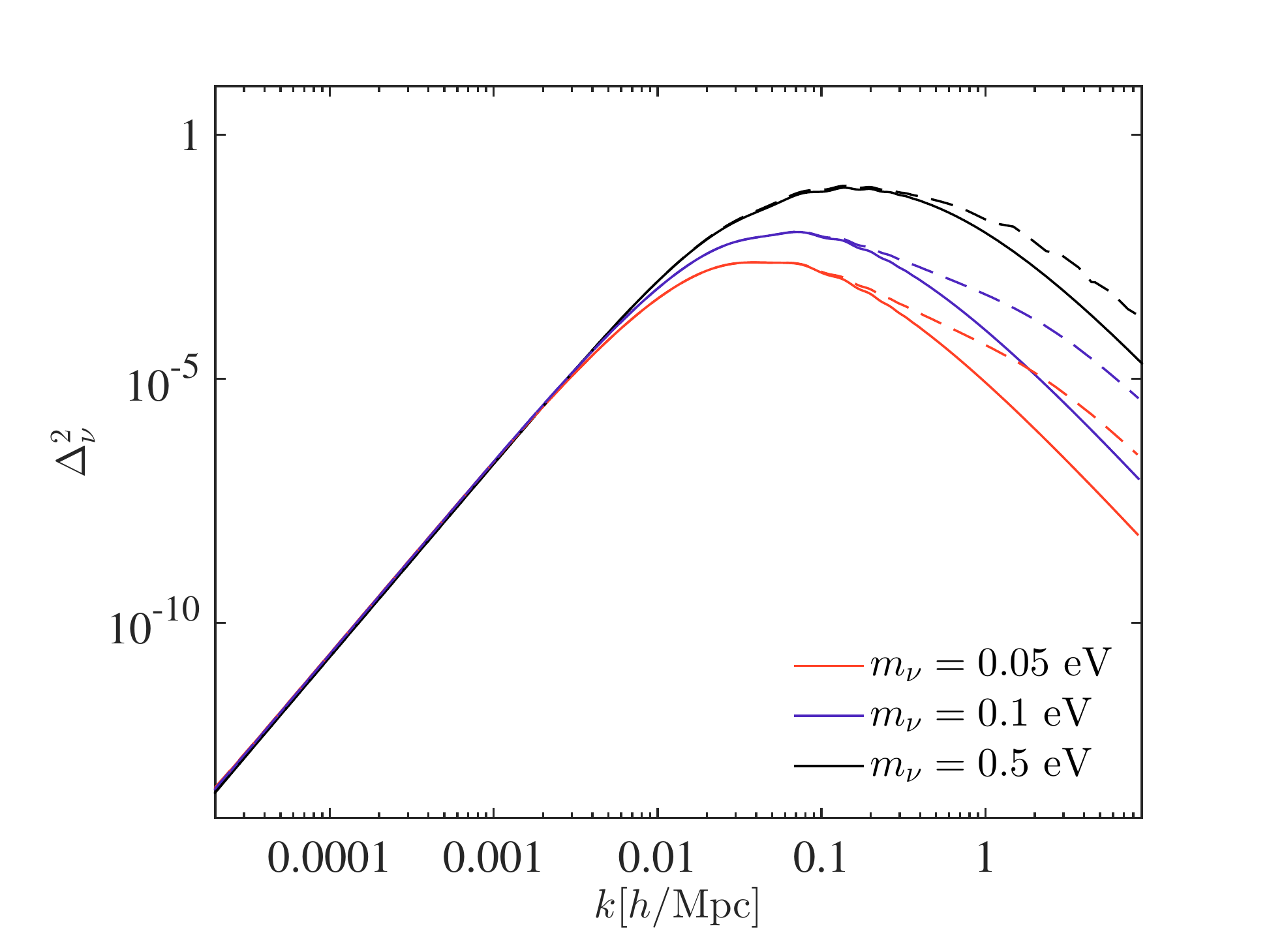}
\end{center}
\caption{The dimensionless neutrino power spectrum ($z=0$). Solid lines represent the linear predictions, dashed lines show the semi-linear model with neutrino overdensities evolving in a non-linear gravitational potential.
\label{fig:abs}}
\end{figure}

The influence of neutrinos in this regime has been studied numerous times in recent years by use of large-scale $N$-body simulations
(see e.g.\ \cite{Brandbyge1,Brandbyge2,Brandbyge3,Brandbyge4,Agarwal:2010mt,Viel:2010bn,Castorina:2013wga,Costanzi:2013bha,Villaescusa-Navarro:2013pva}).
In \cite{Brandbyge2} it was demonstrated that for neutrino masses below $\sim 0.5$ eV the total matter power spectrum can be calculated at a precision of better than $\sim$ 1\% even when the neutrino component is assumed to exactly follow the linear perturbation theory equations (including the linear theory gravitational potential). However, in this case the neutrino power spectrum itself is not reliably calculated because of the absence of the non-linear gravitational source term.
In \cite{AliHaimoud:2012vj} this approach was improved by directly solving the linear neutrino evolution equations on a grid in the $N$-body simulation so that the full gravitational potential could be included.
This has the advantage that the neutrino power spectrum can be accurately calculated as long as neutrino perturbations stay essentially linear on all relevant scales.
In Fig.~\ref{fig:abs} we show the dimensionless neutrino power spectrum calculated using linear theory for different masses. At $z=0$ neutrino perturbations remains linear even for a mass of 0.5 eV, but for masses only slightly higher than that the neutrino density field does enter the non-linear regime. 

We will use a very similar approach, but instead of solving the neutrino evolution equations as a part of the $N$-body simulation we will keep the entire computation in $k$-space.
The non-linear corrections to the gravitational potential are calculated with the \textrm{HALOFIT} version presented in Ref.~\cite{Takahashi:2012em}, including massive neutrinos as in Ref.~\cite{Viel2011}. In order to follow the linear evolution of neutrino perturbations in the non-linear potential, we use a modified version of \textrm{CLASS} \cite{Blas:2011rf}.
The dimensionless neutrino power spectra calculated in CLASS with the full non-linear gravitational potential are shown as dashed lines in Fig.~\ref{fig:abs}.

\subsection{Results}

\begin{figure}[h]
\begin{tabular}{cc}
\includegraphics[width=7cm]{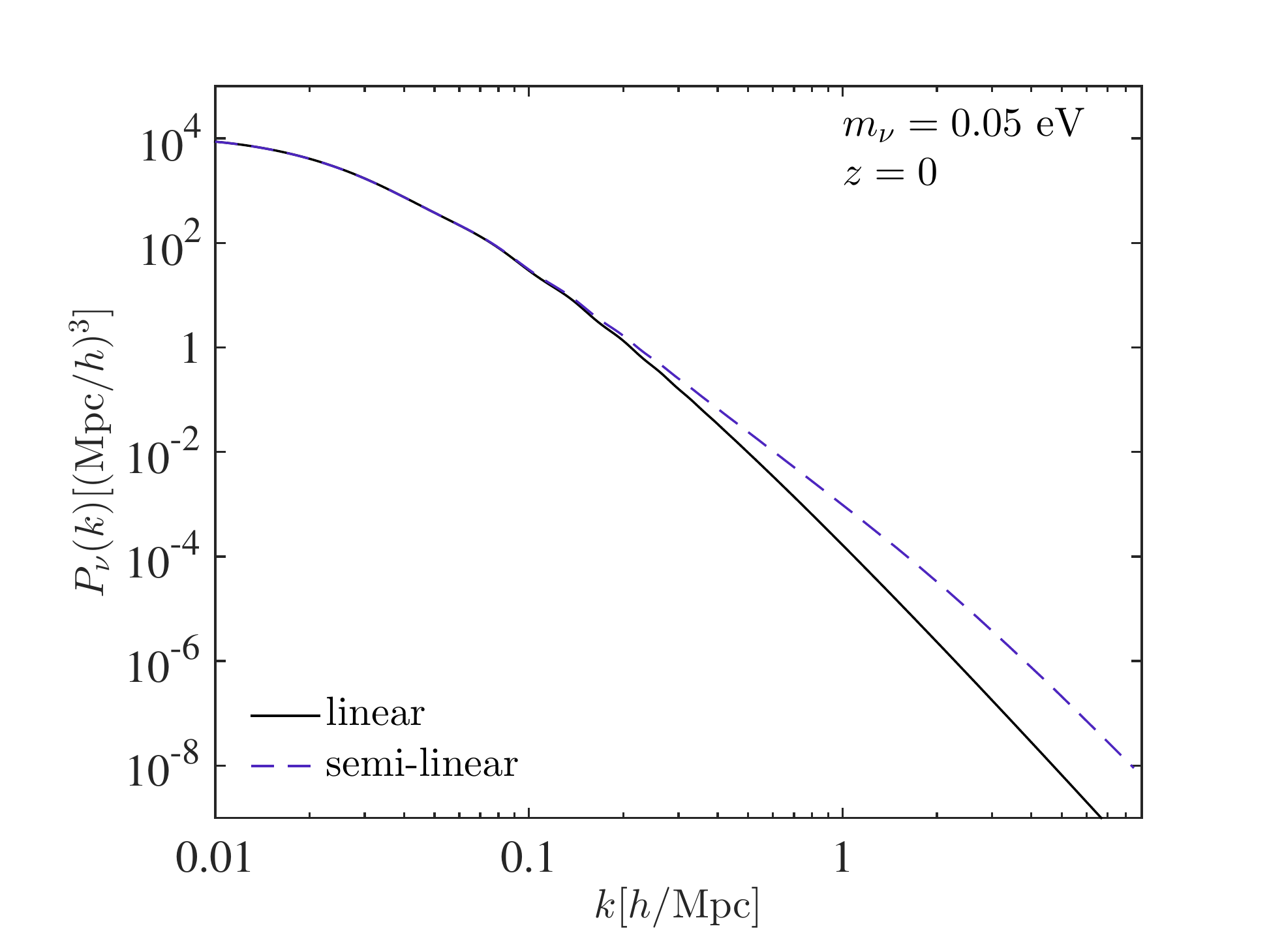}&\includegraphics[width=7cm]{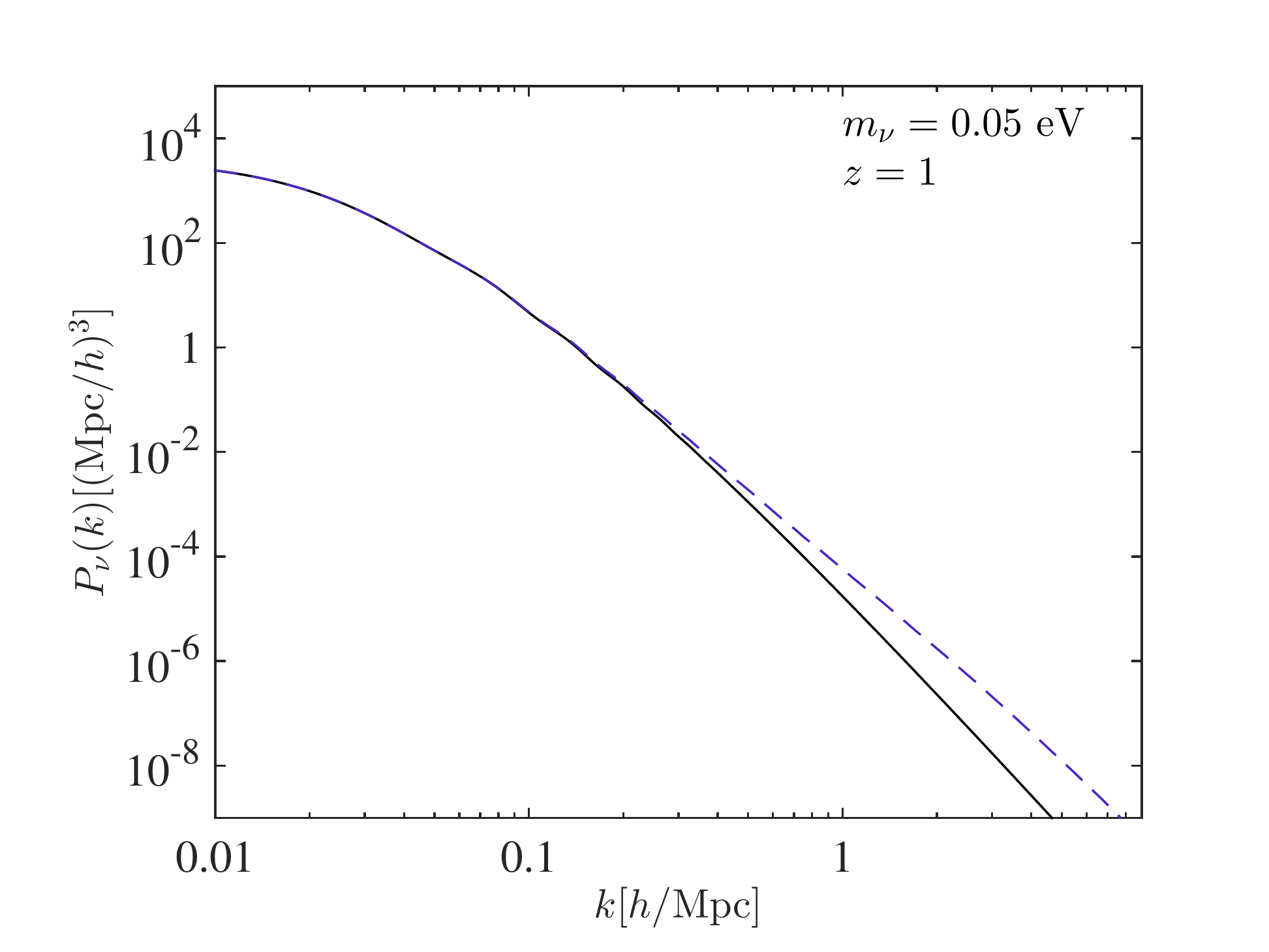}\\
\includegraphics[width=7cm]{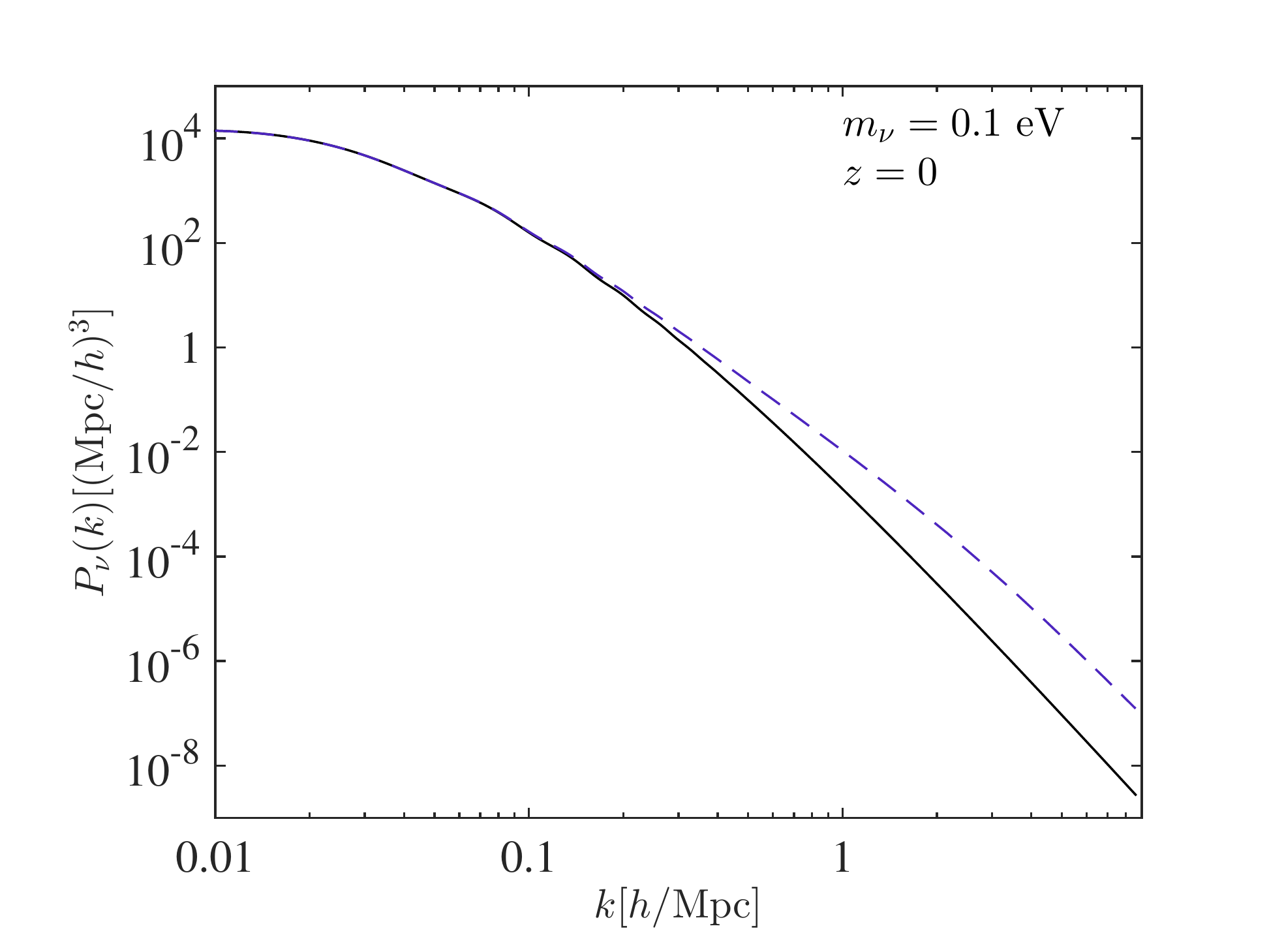}&\includegraphics[width=7cm]{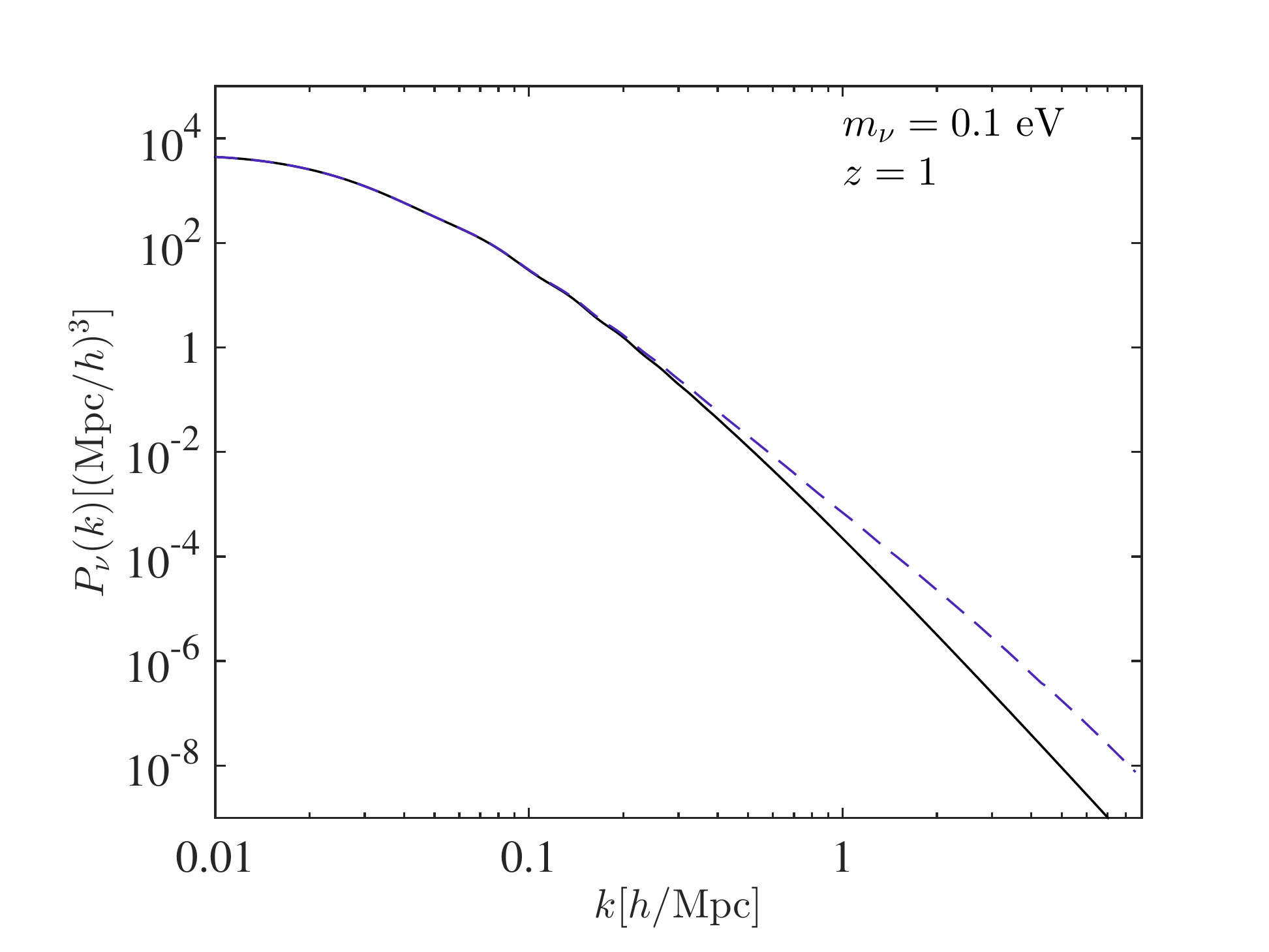}\\
\includegraphics[width=7cm]{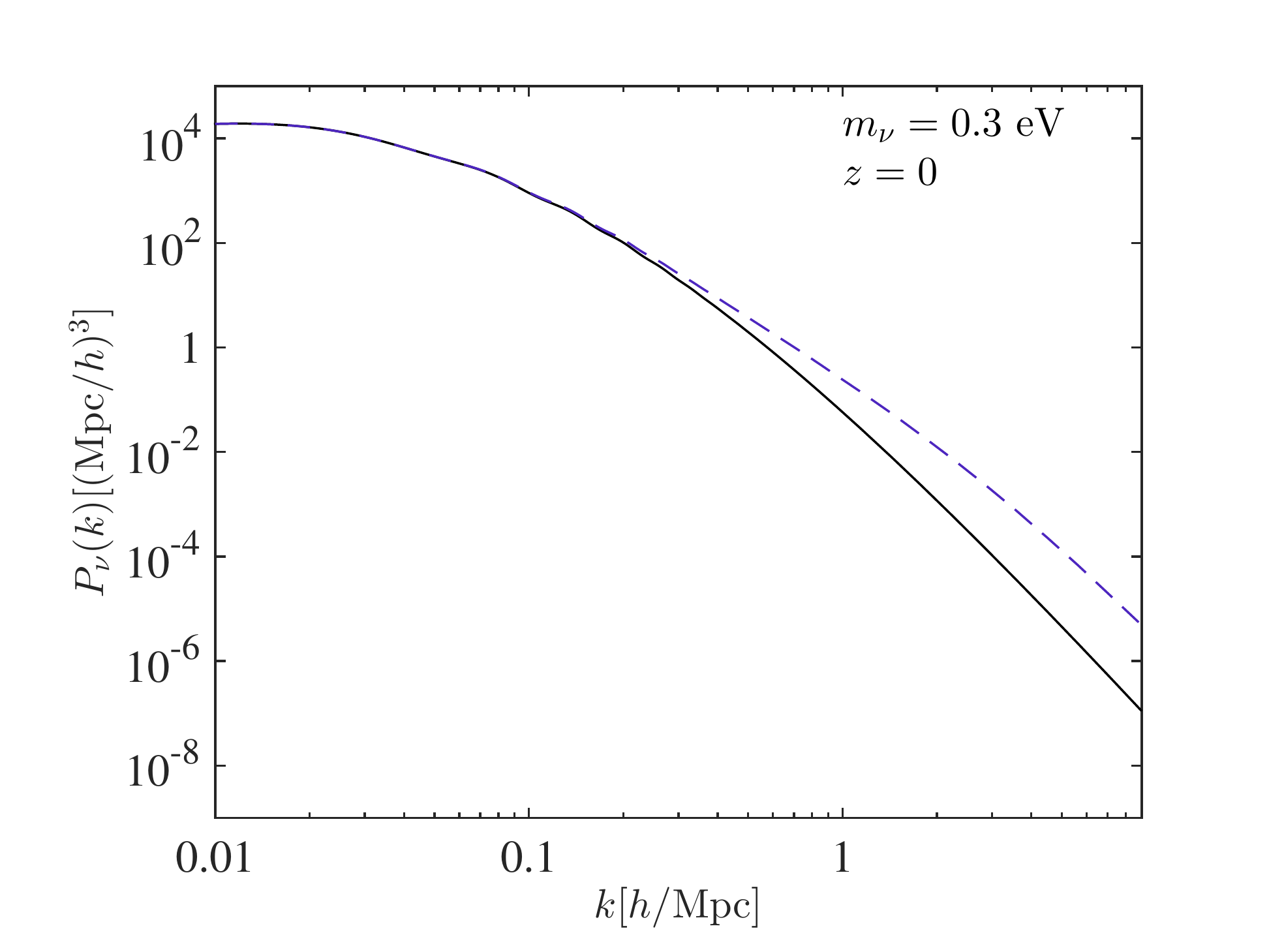}&\includegraphics[width=7cm]{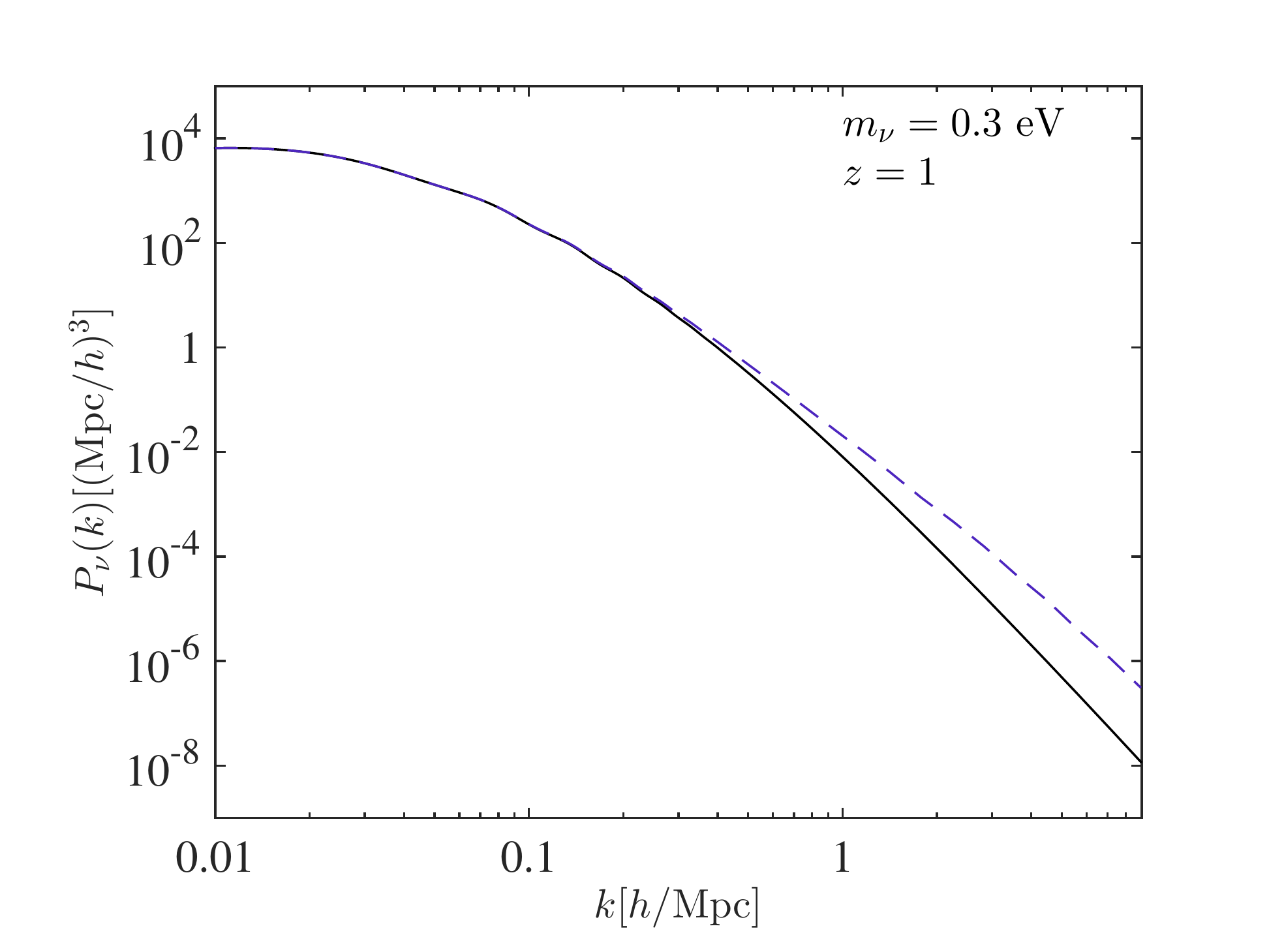}\\
\end{tabular}
\caption{Neutrino component of the matter power spectrum. Solid lines represent the linear predictions, dashed lines show the results of semi-linear approximation. \label{fig:pk}}
\end{figure}

In Fig.~\ref{fig:pk} we show results for the neutrino power spectrum for various masses and redshifts. The full lines are for pure linear theory whereas the dashed lines include the full non-linear potential in the calculation. For a mass of 0.3 eV the increase in neutrino power can be more than a factor of ten relative to linear theory. However, as can also be seen from Fig.~\ref{fig:abs}, in no case do the neutrino perturbations go non-linear, even when the non-linear source term is included.
By combining Fig.~\ref{fig:pk} with Fig.~\ref{fig:ratio} it is also clear that the contribution of this correction to the total matter power spectrum is completely negligible for all masses considered, unlike the effect of neutrinos on the linear power spectrum and the processed effect of this on the non-linear matter power spectrum.

Finally we wish to point out that the correction to the neutrino power spectrum found here is exatly equal to what was found using N-body simulations in
\cite{AliHaimoud:2012vj}, for the same masses and cosmological parameters.

\section{Discussion}

We have demonstrated that the neutrino evolution hierarchy can be solved very accurately even if truncated at $l=2$. Our approximation for the $l=3$ term allowed us to reliably calculate the neutrino power spectrum to better than $\sim 5$\% precision for masses up to $\sim$ 1 eV. The matter power spectrum has a precision of better than $5 \times 10^{-3}$ because of the relatively small direct contribution of neutrinos to this quantity. Finally, our approximation allowed us to calculate the CMB power spectrum with a precision better than $5 \times 10^{-4}$, i.e.\ with a precision far exceeding what is needed in order not to bias parameter estimation from current or future experiments.

The approximation to $\Psi_3$ used here is significantly more precise than previously used once. However, since we keep the momentum dependence in the Boltzmann equations the speed-up factor is less than what can be achieved using the fluid equations. Nevertheless we obtain a speed-up of a factor $2.3$ for standard CLASS settings.

Having obtained a good approximation to $\Psi_3$ we proceeded to study how the neutrino power spectrum can be calculated using the full non-linear gravitational potential, but keeping the entire computation in $k$-space. We obtain results from this technique which are completely consistent with those from $N$-body simulations. However, in our case the neutrino power spectrum can be obtained in a few seconds whereas the $N$-body technique requires far bigger computational resources.

\section*{Acknowledgments}

We wish to thank Julien Lesgourgues and Thomas Tram for valuable discussions and comments on the initial draft.

\clearpage

\bibliographystyle{utcaps}
\providecommand{\href}[2]{#2}\begingroup\raggedright
\endgroup

\end{document}